\documentclass[letterpaper,twocolumn,10pt]{article}

\usepackage{comment,url,algorithm,algorithmic,graphicx,subcaption,relsize,multirow}
\usepackage{amssymb,amsfonts,amsmath,amsthm,amscd,dsfont,mathrsfs,mathtools,nicefrac,stmaryrd,amssymb}
\usepackage{float,psfrag,epsfig,color,xcolor,url,hyperref}
\usepackage{epstopdf,bbm,mathtools,enumitem}
\usepackage{tikz}
\usetikzlibrary{calc, fit, positioning}

\def\balign#1\ealign{\begin{align}#1\end{align}}
\def\baligns#1\ealigns{\begin{align*}#1\end{align*}}
\def\balignat#1\ealign{\begin{alignat}#1\end{alignat}}
\def\balignats#1\ealigns{\begin{alignat*}#1\end{alignat*}}
\def\bitemize#1\eitemize{\begin{itemize}#1\end{itemize}}
\def\benumerate#1\eenumerate{\begin{enumerate}#1\end{enumerate}}

\newenvironment{talign*}
 {\csname align*\endcsname}
 {\endalign}
\newenvironment{talign}
 {\csname align\endcsname}
 {\endalign}

\def\balignst#1\ealignst{\begin{talign*}#1\end{talign*}}
\def\balignt#1\ealignt{\begin{talign}#1\end{talign}}

\let\originalleft\left
\let\originalright\right
\renewcommand{\left}{\mathopen{}\mathclose\bgroup\originalleft}
\renewcommand{\right}{\aftergroup\egroup\originalright}

\def\tinycitep*#1{{\tiny\citep*{#1}}}
\def\tinycitealt*#1{{\tiny\citealt*{#1}}}
\def\tinycite*#1{{\tiny\cite*{#1}}}
\def\smallcitep*#1{{\scriptsize\citep*{#1}}}
\def\smallcitealt*#1{{\scriptsize\citealt*{#1}}}
\def\smallcite*#1{{\scriptsize\cite*{#1}}}

\def\<{\left\langle} 
\def\>{\right\rangle}

\def\norm#1{\left\|{#1}\right\|} 
\def\norm#1{\left\|{#1}\right\|} 
\ifdefined\nonewproofenvironments\else
\ifdefined\ispres\else

\newenvironment{proof-sketch}{\noindent\textbf{Proof Sketch}
  \hspace*{1em}}{\qed\bigskip\\}
\newenvironment{proof-idea}{\noindent\textbf{Proof Idea}
  \hspace*{1em}}{\qed\bigskip\\}
\newenvironment{proof-of-lemma}[1][{}]{\noindent\textbf{Proof of Lemma {#1}}
  \hspace*{1em}}{\qed\\}
\newenvironment{proof-of-theorem}[1][{}]{\noindent\textbf{Proof of Theorem {#1}}
  \hspace*{1em}}{\qed\\}
\newenvironment{proof-attempt}{\noindent\textbf{Proof Attempt}
  \hspace*{1em}}{\qed\bigskip\\}

\fi

\fi
\makeatletter
\@addtoreset{equation}{section}
\makeatother

\hypersetup{
  colorlinks,
  linkcolor={red!50!black},
  citecolor={blue!50!black},
  urlcolor={blue!80!black}
}

\usepackage{url}
\usepackage{todonotes}

\usepackage{usenix2019_v3,epsfig,endnotes}
\usepackage{booktabs}
\usepackage{tcolorbox}
\begin{document}

\date{}

\title{On the Exploitability of Audio Machine Learning Pipelines to Surreptitious Adversarial Examples}

\author{Adelin Travers$^{\dag \ddag}$, Lorna Licollari$^{\dag}$, Guanghan Wang$^{\dag}$, Varun Chandrasekaran$^{*}$, Adam Dziedzic$^{\dag \ddag}$, \\ David Lie$^{\dag}$, Nicolas Papernot$^{\dag \ddag}$ \\ $\ddag$Vector Institute,$\dag$University of Toronto,$^{*}$University of Wisconsin-Madison }

\maketitle

\begin{abstract}

Machine learning (ML) models are known to be vulnerable to adversarial examples. Applications of ML to voice biometrics authentication are no exception. Yet, the implications of audio adversarial examples on these real-world systems remain poorly understood given that most research targets limited defenders who can only listen to the audio samples. Conflating \textit{detectability} of an attack with human \textit{perceptibility}, research has focused on methods that aim to produce imperceptible adversarial examples which humans cannot distinguish from the corresponding benign samples. We argue that this perspective is coarse for two reasons: 1. Imperceptibility is impossible to verify; it would require an experimental process that encompasses variations in listener training, equipment, volume, ear sensitivity, types of background noise etc, and 2. It disregards pipeline-based detection clues that realistic defenders leverage. This results in adversarial examples that are ineffective in the presence of \textit{knowledgeable defenders}. Thus, an adversary only needs an audio sample to be \textit{plausible} to a human. We thus introduce \textit{surreptitious adversarial examples}, a new class of attacks that evades \textit{both human and pipeline controls}. In the white-box setting, we instantiate this class with a joint, multi-stage optimization attack. Using an Amazon Mechanical Turk user study, we show that this attack produces audio samples that are more surreptitious than previous attacks that aim solely for imperceptibility. Lastly we show that surreptitious adversarial examples are challenging to develop in the black-box setting.

\end{abstract}

\section{Introduction}
\label{sec:intro}

Machine Learning (ML) based voice biometrics have increasingly been implemented by financial institutions to complement or even fully replace standard authentication~\cite{noauthor_hsbc_2016,corera_banks_2016,sheets_biometrics_2019,timem_utilizing_2015}. In our work, we examine the security of voice biometrics in the light of adversarial examples~\cite{biggio_evasion_2013,szegedy_intriguing_2013,abdullah_sok_2020,carlini_audio_2018}. Defending against adversarial examples remains a largely open problem; model-level ML defenses can be evaded by accommodating existing optimization frameworks in a strategy known as \textit{adaptive attacks}~\cite{carlini_towards_2017,tramer_adaptive_2020}, putting adversaries at an advantageous position. At the level of a  system, this view is largely due to two \textit{threat model simplifications} the community has made to define adversarial examples.  

\textit{First}, the defenders considered in prior work usually have very limited capabilities, \textit{e.g.}, the defender may only listen to the audio samples. Hence, prior work focused on the coarse concept of \textit{human perceptibility}, \textit{i.e.,} an adversarial example cannot be distinguished from the corresponding benign sample, or is obfuscated by specific signals. As this goal is hard to formalize, we show through a human study that state-of-the-art attacks that target human perceptibility fail to fool lightly trained humans (a proxy for knowledgeable defenders), suggesting that true imperceptibility is \textit{hard to achieve}. \textit{Secondly}, prior work makes a simplifying assumption that the adversary is able to modify the model's inputs \textit{directly} and does not have to consider any constraints placed on these inputs by existing \textit{pre-processing} steps preceding the model (all of which are collectively called a \textit{pipeline}). Thus, prior attacks neglect checks at different steps of this pipeline (that are simpler to implement), like checking for spectrogram artifacts. 
To summarize, on the one hand human imperceptibility is too strict of a requirement whereas on the other hand other aspects of the pipeline are often ignored. 

In this paper, we introduce a more comprehensive attack model for ML in audio which changes the objective an attacker would optimize. We first show that \textit{audio plausibility}--the property where a human listener believes that an (adversarial) audio sample they hear \textit{could have} originated from a benign source--alone is needed to fool humans. While this may be a weaker requirement than imperceptibility, which requires that the adversarial example be indistinguishable from a benign sample, it is a \textit{more realistic} requirement as it assumes the human is aware of and looking for adversarial perturbation. 

We argue that the most compelling attacks form a \textit{new attack class} that are not only human plausible but also not abnormal to pipeline based controls. We term these strong attacks \textit{surreptitious} attacks. We show that attacking a single step of the pipeline (\textit{e.g.}, a pre-processing step, or the input of the model) is insufficient to build a surreptitious attack (as it can be detected by constraints at other pipeline stages). 
To improve both current defenses and future attack evaluation, we identify constraints stemming from the pipeline for common deep learning-based automated speaker identification (ASI) pipelines. As a consequence, our adversary model also takes into account detectability through these pipeline constraints. 

In the white-box case, we instantiate the surreptitious class on the LibriSpeech dataset~\cite{panayotov_librispeech_2015} by \textit{jointly optimizing} a standard attack objective and a novel surreptitious objective formulated from pipeline intermediate features. Despite larger digital audio perturbations than prior work, an IRB approved Amazon Mechanical Turk (MTurk) study we conducted finds that humans find the audio samples generated by our attacks similarly or more plausible than those of previous attack techniques. 
Lastly, we show that achieving surreptitiousness is challenging in the pipeline-level black-box setting. Since the objective leverages intermediate features, direct black-box approaches like finite differences (or gradient free optimization techniques) cannot be used. Contrary to model-level transfer attacks used in prior work, \textit{pipeline-level} surreptitious attacks have the added troublesome requirement that intermediate features of the end-to-end differentiable approximation and the target pipeline be sufficiently close. 

Our contributions can be summarized as follows:
 \begin{itemize}

\item We show that focusing on the perceptibility of perturbations does not lead to an undetectable attack, because both humans and algorithmic heuristics may detect perturbations produced by attack algorithms searching for  imperceptible perturbations.

\item We introduce a new class of attacks, surreptitious adversarial examples, which are plausible to humans (\textit{e.g.}, could be benign samples) and respect algorithmic detection heuristics (\textit{i.e.,} constraints) in place at different stages of the audio ML pipeline. 

\item We introduce a white-box attack which achieves surreptitiousness by jointly optimizing objectives defined at the level of the model and relevant stages of the pipeline. We show that searching for surreptitious adversarial examples allows the attacker to trade-off the use of its perturbation budget to effectively bypass (human or algorithmic) defense mechanisms put in place by the defender at these different pipeline stages. We validate the human plausibility of the attack through an Amazon Mechanical Turk study. 

\item We devise a black-box attack approximating the intermediate stages of the pipeline are unavailable to the adversary, and evaluate its effectiveness, finding it less effective for surreptitiousness than white box attacks.
 \end{itemize}

One of the conclusions of our work is that limitations in the  threat model previously considered  led to an overly pessimistic perspective on  ML robustness to adversaries in audio. 
Instead, our work  takes a system perspective. We provide evidence that reasoning about robustness in an end-to-end fashion rather than at the level of an isolated model provides a more realistic assessment of robustness. This enables concrete progress from both the defender's (see Section~\ref{sec:beyond}) and adversary's (see Sections~\ref{sec:wbox} and~\ref{sec:snes}) points of view. 
\begin{figure}[H]
    \centering
    \resizebox{\linewidth}{!}{
    \begin{tikzpicture}
\Large

\makeatletter
\tikzset{
	fitting node/.style={
		inner sep=0pt,
		fill=none,
		draw=none,
		reset transform,
		fit={(\pgf@pathminx,\pgf@pathminy) (\pgf@pathmaxx,\pgf@pathmaxy)}
	},
	reset transform/.code={\pgftransformreset}
}
\makeatother

\tikzset{
	every rectangle node/.style={
		align=center
	},
	sine/.style={
	path picture={
		\draw[x=1.57ex, y=1ex] (0,0) sin (1,1) cos (2,0) sin (3,-1) cos (4,0) (0,1) cos (1,0) sin (2,-1) cos (3,0) sin (4,1);
	}
},
	fit margins/.style={/tikz/afit/.cd,#1,
		/tikz/.cd,
		inner xsep=\pgfkeysvalueof{/tikz/afit/left}+\pgfkeysvalueof{/tikz/afit/right},
		inner ysep=\pgfkeysvalueof{/tikz/afit/top}+\pgfkeysvalueof{/tikz/afit/bottom},
		xshift=-\pgfkeysvalueof{/tikz/afit/left}+\pgfkeysvalueof{/tikz/afit/right},
		yshift=-\pgfkeysvalueof{/tikz/afit/bottom}+\pgfkeysvalueof{/tikz/afit/top}},
		afit/.cd,left/.initial=2pt,right/.initial=2pt,bottom/.initial=2pt,top/.initial=2pt
}


\path[draw, x=1.57ex, y=1ex, xshift=-3.5cm] (0,0) sin (1,1) cos (2,0) sin (3,-1) cos (4, 0) node[fitting node, label={Audio\\Wave}] (W){};

\path[minimum height=2.5cm, minimum width=2.5cm, rounded corners]
	node (AS) [draw] at (0, 0) {Acquistion\\System}
	node (ADC) [right=1.5cm of AS.east, draw, rectangle,  draw] at (0, 0) {Analog to \\Digital\\Converstion}
	node (AP) [draw, fit=(AS)(ADC), fit margins={top=15pt}] {}
	
	node (SP) [right=1cm of ADC.east, draw]  {Sound\\Processing}
	node (SpP) [right=0.5cm of SP.east, draw] {Spectral\\Processing}
	node (DP) [draw, fit=(SP)(SpP), fit margins={top=15pt}] {}
	
	node (L0) [right=1cm of SpP.east, draw, minimum width=0.5cm, ] {$L_0$}
	node (Ldot) [right=0.2cm of L0.east, minimum width=0.5cm] {...}
	node (CL) [right=0.5cm of Ldot.east, minimum width=0.5cm, draw] {$L_{-1}$}
	node (M) [draw, fit=(L0)(CL), fit margins={top=15pt}] {};

\path[anchor=north, inner sep=7pt]
	node at (AP.north) {Analog Processing}
	node at (DP.north) {Digital Processing}
	node at (M.north)  {Model};
	
\newcommand{\raisesep}{3cm}
\newcommand{\dropsep}{2cm}
\node (S0) at ($(Ldot.east)!0.5!(CL.west) - (0, \dropsep)$) {$S_0$};
\node (S1) at ($(SpP.east)!0.5!(L0.west) + (0, \raisesep)$) {$S_1$};
\node (S2) at ($(SP.east)!0.5!(SpP.west) - (0, \dropsep)$) {$S_2$};
\node (S3) at ($(ADC.east)!0.5!(SP.west) + (0, \raisesep)$) {$S_3$};
\node (S4) at ($(W.east)!0.5!(AS.west) - (0, \dropsep)$) {$S_4$};

\draw [->] ($(W.east) + (0.1cm, 0)$) to (AS.west);
\draw [->] (AS.east) to (ADC.west);
\draw [->] (ADC.east) to (SP.west);
\draw [->] (SP.east) to (SpP.west);
\draw [->] (SpP.east) to (L0.west);
\draw [->] (L0.east) to (Ldot.west);
\draw [->] (Ldot.east) to (CL.west);
\draw [->] (S0) to ($(Ldot.east)!0.5!(CL.west)$);
\draw [->] (S1) to ($(SpP.east)!0.5!(L0.west)$);
\draw [->] (S2) to ($(SP.east)!0.5!(SpP.west)$);
\draw [->] (S3) to ($(ADC.east)!0.5!(SP.west)$);
\draw [->] (S4) to ($(W.east)!0.5!(AS.west)$);

\end{tikzpicture}
    }
    \caption{Audio processing pipeline, with stages where adversarial perturbations can be injected.\vspace{-5mm}}
    \label{fig:pipeline}
\end{figure}
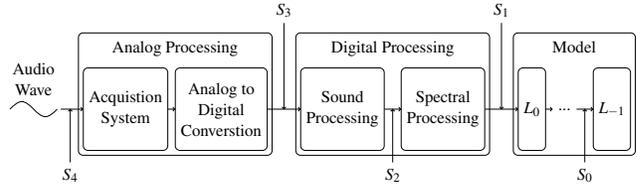

\section{Audio Processing Fundamentals}
\label{sec:background}

An analog waveform or audio sample is considered as an input $x$ in a space $X$. A \textit{pipeline} is conceptually a mapping,  including the classifier itself, between an input and a label $t$ in a space $T$. Formally, we define a pipeline as the sequence of operations: $x\xrightarrow{p_1} s_1 \cdots  \xrightarrow{p_N} s_N \xrightarrow{m} t$, where $p_i$ denotes a function for \textit{pre-processing}. $m$ denotes a \textit{classification function} for the pre-processed input. This abstract view is embodied by the audio processing pipeline schematic in Figure~\ref{fig:pipeline}. We highlight important steps and provide rudiments to understand the role of each of these stages below; a more detailed treatment of audio processing is given in Appendix~\ref{app:sec:preproc}. Note that generic audio processing pipelines may also include additional steps, \textit{e.g.}, context tracking or post-processing in ASR. Hence we number the steps from the classification component as it is the only non-optional element common to all pipeline types.

\begin{description}
\item[\textbf{Stage 4}:] (\textit{Analog audio stage}). This is the analog wave $x$ before any pre-processing. The analog waveform is first acquired and converted to a digital representation by sampling and quantization by the analog processing component (see Appendix \ref{app:subsec:A2D}).

\item[\textbf{Stage 3}:] (\textit{Digital audio stage}). The input to the digital processing stage of the pipeline is the sound wave after encoding (see Appendix~\ref{app:subsec:digital_processing}).

\item[\textbf{Stage 2}:] (\textit{Feature stage}).  The pipeline produces a feature representation of the input for each of its $1 \leq i < N$ pre-processing steps, before it produces the input to the model (see Appendix~\ref{app:subsec:Fourier}). 

\item[\textbf{Stage 1}:] (\textit{Model stage}). This is the input to the model $m$, \textit{e.g.}, the spectrogram corresponding to the audio sample. Depending on the preprocessing steps, different types of models, typically Convolutional Neural Nets (CNNs), will be used with these inputs as shown below.

\item[\textbf{Stage 0}:] (\textit{Classification stage}). This is the final classification layer, such as the softmax layer. 
\end{description}

Not all models are compatible with all possible processing steps. Hence choosing a model or a set of preprocessing steps fully determines a pipeline. In our work, we study three classes of pipeline. We present them in order of decreasing reliance on (analog and digital) pre-processing (see Appendix~\ref{app:sec:architectures} for details about model architectures): 

\begin{itemize}
\itemsep0em
\item \textbf{Audio-based Pipelines (\texttt{ABP}).} The earliest audio-specific pipelines operated on hand-crafted features such as MFCCs~\cite{davis_comparison_nodate}. We consider X-vector~\cite{snyder_x-vectors_2018}, which combines  MFCC  based pre-processing with time delay neural networks (TDNNs), as a representative example. 

\item \textbf{Spectrogram-based Pipelines (\texttt{SBP}).} Hand engineering features for audio is challenging and it is not clear if ML systems (including DNNs) should be designed in an anthropomorphic fashion. New research~\cite{nagrani_voxceleb_2017,Nagrani20} focused on building audio processing techniques directly atop \textit{spectrograms}. This is particularly appealing since the spectrogram can be treated as an image, and architectures from computer vision adapted to the audio domain. We choose Vggvox~\cite{nagrani_voxceleb_2017}, a modified 2D CNN, as a representative example of this type of pipeline. 

\item \textbf{End-to-end DNN-based Pipelines (\texttt{DBP}).} End-to-end audio models~\cite{palaz_analysis_nodate,ravanelli_speaker_2019} are a new alternative to the \texttt{ABP} and  \texttt{SBP} approaches. Instead of manually performing pre-processing (\textit{i.e.,} computing the STFT~\cite{sturmel_signal_2011} and Mel scale~\cite{davis_comparison_nodate}), layers of an end-to-end 1D CNN model embed these computations. The model learns how to approximately compute these operations.\footnote{\url{https://github.com/mravanelli/SincNet/issues/74}} We utilize SincNet~\cite{ravanelli_speaker_2019} as a representative example.
\end{itemize}

\section{Beyond Human Perceptibility}
\label{sec:beyond}

The community has focused its efforts on human perceptibility of adversarial examples by either creating \textit{inaudible} perturbations~\cite{abdullah_hear_2019,taori_targeted_2019,carlini_audio_2018} or obfuscating voice commands with high power noise~\cite{abdullah_practical_2019,carlini_hidden_2016} or songs~\cite{devil,commandersong}.
Our work is based on the key observation that, albeit never explicit in prior work, \textit{the imperceptibility argument serves as a proxy for the property that the attack is hard to detect, if only to a human observer}. 
This is problematic for two reasons:

\begin{description}
\item[\textbf{Coarse Concept (P1)}:] It does not distinguish between experts and layman sensitivity to attacks, nor the quality of hardware and software they have access to. In our (the authors) experience listening to attack samples, even with fixed equipment, variables like volume or concentration impacted our capability to perceive generated samples. 

\item[\textbf{Pipeline detection (P2)}:] It does not consider clues that are \textit{not} in the human perceptible domain of audio. Such clues, \textit{e.g.}, spectral anomalies, are available to strong defenders.
\end{description}

\subsection{Plausibility vs. Imperceptibility}
\label{subsubsec:overkill}

The coarseness (\textbf{P1}) leads to \textit{attack overkill}: the attacker does not \textit{need} to ensure that the adversarial example is indistinguishable from the original sample; she only has to ensure that she is not trivially detected by a human listener. In particular, we say that (i) a sample is \textit{imperceptible} iff it cannot be distinguished by human perception (across a variety of listening conditions) from the benign sample used to generate it, and (ii) a sample is \textit{plausible} iff humans trained to recognize abnormal inputs for the application at hand think it could have been obtained by a natural channel for this application (\textit{e.g.}, a benign but noisy phone channel). Observe that (i) an imperceptible adversarial example is necessarily plausible but the converse is not true, and (ii) an obfuscated attack, \textit{i.e.,} covering a malicious command with a signal such as noise or songs~\cite{abdullah_practical_2019,commandersong}, may not be plausible if the obfuscating signal is \textit{abnormal} for the target application. Lastly, observe that the definition of imperceptibility, as used in prior work, is strong and hard if not impossible to prove since it would require to be independent of hardware, training, specific application at hand, volume or any other parameters that may influence distinguishability from the benign sample. In contrast, our definition of plausibility is tied to the specific problem at hand and is easier to validate experimentally.

This suggests that imperceptibility is often a \textit{stronger requirement} than what is actually needed (\textit{i.e.,} plausibility); by targeting imperceptibility, an adversary is using overly stringent constraints when searching for a perturbation. This is in stark constraint with more realistic adversaries which are work adverse~\cite{allodi_workaverse_2021}. Since the adversary is attempting to solve a more difficult problem, they may also expend unnecessary computational cycles in an iterative attack like projected gradient descent (PGD)~\cite{kurakin_adversarial_2017,madry_towards_2019} where the simpler and cheaper fast gradient sign method (FGSM)~\cite{szegedy_intriguing_2013} might suffice. In summary, unless the added computational expense helps reduce attack detectability, targeting imperceptibility makes the attack cost inefficient.

We also observe that very little training is sufficient to equip defenders with abilities to identify \textit{abnormal} adversarial examples (those that are masked with audio which is not relevant to the application being fooled). We illustrate this with a technique proposed by Yuan \textit{et al.}~\cite{commandersong} and later refined by Chen \textit{et al.}~\cite{devil}, where a song is used to mask a malicious command. The command is inserted in the song using gradient descent\footnote{This method demands both commitment and expertise as the authors had to "manually analyze the source code of Kaldi (about 301,636 lines of shell scripts and 238,107 C++ SLOC)" (\textit{sic.}~\cite{commandersong}).}. The authors test the identifiability of their attack on \textit{non-experts}. While the authors report that the commands are not recognized, a survey they conducted suggests that more than 90\% of end-users identify speech in the Yuan \textit{et al.}~\cite{commandersong} attack and more than 60\% identify noise in the Chen \textit{et al.} attack~\cite{devil} (both of which are deemed abnormal). 

Note that voice authentication in several sectors such as finance have \textit{proactive} defenders who identify threats~\cite{noauthor_proactive_nodate,noauthor_threat_2020,noauthor_threat_nodate,noauthor_what_nodate}. Trained personnel investigating a (fraudulent) authentication attempt to a financial system, for instance after a customer reported suspicious account activity, would notice a song being played rather than a user speaking\footnote{This is even more likely in the newer version of the attack, which is aptly named "Devil's whisper" because the resulting samples have the distinctive "waterfall" chirp of audio adversarial examples.}. In a study presented in \S~\ref{subsec:mturk}, we confirm that even minimally trained MTurkers can identify the refined attack~\cite{devil} as abnormal in 93.7\% of cases simply by hearing the samples. This is despite extensive effort being placed on minimizing the difference between malicious and benign samples.

We conclude by re-iterating that imperceptibility is both \textit{unnecessary} (because plausibility is sufficient) and \textit{hard to achieve} (because it is still detectable with minimal training) for the attacker despite its associated costs. Put another way, we argue that perceptibly, more specifically the audibility argument for ASI, is a \textit{red herring} rooted in insufficient defender capabilities compared with realistic defenders. It leads attackers to consider the wrong constraints when formulating their optimization problem.  

\subsection{Pipeline-induced Constraints}
\label{subsec:constraints}

Recall that audio differs greatly from domains like computer vision where the model is directly fed with the raw input (\textit{e.g.}, image pixels). Instead, audio-based and spectrogram-based pipelines rely on several pre-processing stages ($S_i$ in Figure~\ref{fig:pipeline}) which we described when introducing a typical audio pipeline (\textit{e.g.}, \texttt{ABP} and \texttt{SBP} in \S~\ref{sec:background}). Each of these stages constitutes a possible \textit{insertion point} for the adversary if she is able to modify the pipeline at this stage;\footnote{
To see how an adversary could modify the pipeline's internal stages, consider the \textit{client-server} architecture. In audio processing, the server may have dedicated ML accelerators and hardware (TPUs or GPUs) reserved for matrix algebra computation~\cite{noauthor_limesdr_nodate}. In this setting, $x\xrightarrow{p_i}s_i$ is part of the computation performed on the client-side (for $i \in [N]$), and $s_i\xrightarrow{m}t$ is part of the computation performed on the server-side. ASRs or ASIs may be directly accessed through APIs, or over the network~\cite{etsi_etsi_nodate,fusakawa_speaker_2016,chao_text_2020,shagalov_client-server_2016,han_method_2019,cho_speech_2020}. A low stage attack can be performed by crafting a malicious client or simply with non-transparent proxies such as Burp Suite or OWASP ZAP~\cite{burp,ZAP} if the application does not enforce certificate pinning.} similar discussions can be found in prior work~\cite{abdullah_sok_2020,pierazzi_intriguing_nodate,1400-1700_isoiec_2020}.  
Conversely, a defender may use the different stages of the pipeline to detect inconsistencies in adversarial examples; the defender is not limited to \textit{only} those stages that are relevant to imperceptible audio perturbations. 

In audio, the few prior works that consider pre-processing either attack the pre-processing itself without considering the specifics of the model~\cite{abdullah_hear_2019,abdullah_practical_2019}, or treat it as an extension of the model by re-implementing it in a differentiable manner~\cite{carlini_audio_2018}. We show that: (i) even if the adversary has the \textit{strong capability} to insert a perturbed input at different stages of the pipeline, (ii) \textit{various} constraints will be violated in stages \textit{preceding} the insertion point. To do so, we give three examples of simple algorithmic analyses of intermediate pipeline stages allowing a defender to detect five state-of-the-art attacks~\cite{abdullah_practical_2019,koerich_cross-representation_2020,kreuk_fooling_2018,marras_adversarial_2019,taori_targeted_2019} on human perception covering all major types identified in the SoK from Abdullah \textit{et al.}~\cite{abdullah_sok_2020} (see \S~\ref{subsubsec:overkill} and Appendix~\ref{app:sec:kenan} for the other attacks).

\subsubsection{Waveform Reconstruction (Stage 4 attacks)}
\label{subsubsec:noisy_attack}

Consider the noisy attack on automated speaker recognition (ASR) proposed in Abdullah \textit{et al.}~\cite{abdullah_practical_2019}. This attack obfuscates malicious audio commands with unintelligible noise. The noise is removed by the pipeline's low-pass filter, leaving the obfuscated sample spectrogram close to the spectrogram of the non-obfuscated malicious command. 

In our formalism: $x_{\text{adv}}+n\xrightarrow{\text{filter}}x_{\text{adv}}\xrightarrow{p}s_{\text{adv}}$ (note the absence of the noise $n$ after the filter). However, a defender could deploy a countermeasure based on an analysis of the malicious spectrogram. The defender would invert the spectrogram (which does not contains noise) $s_{\text{adv}}\xrightarrow{\widetilde{p^{-1}}}\widetilde{x_{\text{adv}}}$ using the (approximate) mapping $\widetilde{p^{-1}}$ from spectrogram-to-signal reconstruction like the Griffin-Lim algorithm~\cite{sturmel_signal_2011}. The defender could then listen to the reconstructed sample $\widetilde{x_{\text{adv}}}$ and flag it as malicious in the likely event that the noise $n$ injected by the attack is now absent from the reconstructed sample, i.e., the attack would then be de-obfuscated. \footnote{Furthermore, the attack also makes several limiting assumptions because it is implicitly a replay attack. These assumptions increase the adversary's cost as illustrated by the following two examples: (i) the attacker must record a legitimate user ahead of time, and (ii) the system must already be vulnerable to replay attacks which is not always the case given that commercial biometric ASI are already often hardened against replay attacks~\cite{1400-1700_isoiec_2020-1,1400-1700_isoiec_2020}.} 
We did not test this hypothesis experimentally, but observe that the inversion process is a bijection under a set of conditions set forth in Sturmel \& Daudet~\cite{sturmel_signal_2011} and that the adversarial command is in this case not an imperceptible adversarial example but rather a full sentence representing a malicious command.

 \subsubsection{Hermitian Symmetry (Stage 1 attacks)}
\label{subsubsec:pipeline_control}

Since a real number is its self complex conjugate, Fourier Transforms on real-valued, by opposition to complex-valued, signals exhibit a symmetry, the \textit{Hermitian symmetry}, where the real and complex coefficients of the (complex) transform are the same. All physical waveforms are real-valued. This symmetry allows us to compute only half of the components of the DFT for real signals. However, this optimisation is not always performed, for example, in Vggvox~\cite{nagrani_voxceleb_2017}, where the output spectrograms contain the redundant information and exhibit the Hermitian (\textit{a.k.a}. conjugate) symmetry.

We experimentally verified this hypothesis by reviewing raw \texttt{numpy} arrays for spectrogram level adversarial examples (see \S~\ref{sec:strawman} for the detailed experimental setup). We observe that neither stage 1 attacks like FGSM nor PGD respect the Hermitian symmetry. On the contrary, this symmetry is naturally enforced by attacks on the digital audio. By algorithmically verifying if the spectrogram is indeed symmetrical, the defender can detect spectrograms that have been tampered with.  From a developmental perspective, this is a simple yet very effective constraint to enforce. In our implementation, this involves \textit{one line of code}, and it \textit{detects all} sprectrum-based stage 1 attacks because they directly manipulate the spectrogram~\cite{koerich_cross-representation_2020,kreuk_fooling_2018,marras_adversarial_2019}. In Appendix~\ref{app:sec:adaptive_symmetry}, we further show that while an adaptive adversary can evade this control, this requires a larger perturbation and computational expense.

 \subsubsection{Nyquist Frequency (Stage 3 attacks)}

With limited knowledge of the pipeline, an attacker focusing on imperceptibility may insert or even constrain attack frequencies between the practical and theoretical Nyquist frequency (as done by Taori\textit{ et al.}~\cite{taori_targeted_2019}). By Shannon's theorem, the maximum frequency that a sampled signal can represent \textit{i.e.,} the \textit{theoretical} Nyquist frequency, is half its sample rate. Broadband is sampled at 16kHz and narrowband at 8kHz with 8kHz and 4kHz the respective theoretical Nyquist frequencies. A low-pass filter in the analog to digital converter will remove frequencies above the Nyquist frequency to prevent aliasing. For real applications, the cut-off frequency is lower than the theoretical Nyquist frequency~\cite{noauthor_g7111_2020, noauthor_g729_2020,noauthor_g726_2020}. The resulting \textit{practical} Nyquist frequency is typically 7kHz and 3.5kHz for broadband and narrowband. Frequencies in the 7 to 8kHz and 3.5 to 4kHz bands should be strongly attenuated.  

A defender can monitor unusual power in the practical or theoretical Nyquist frequency range or simply enforce a low-pass filter at reception on the server side. 
We test this second approach by reproducing the genetic algorithm ASR attack by Taori \textit{et al.}~\cite{taori_targeted_2019} on ASI, targeting speaker misclassification instead of mistranscription. Since the genetic algorithm employed does not minimize the perturbation size, the attack would be very perceptible by default. As noted by the authors~\cite{taori_targeted_2019}, this attack needs to constrain the perturbation in the 7 to 8kHz band to reduce perceptibility as humans are less sensitive to high frequencies. As gradient-free methods, genetic algorithms are inherently slow to converge and computationally costly. Hence, we limit ourselves to a 150 samples chosen at random from the 1000 samples used in experiments of \S\ref{sec:wbox} and run the attack for 10 to 1000 iterations. Once the samples are generated, they are passed through a low frequency filter before being fed to \texttt{SBP} to simulate the practical Nyquist frequency constraint and test 2 cut-off configurations: (i) 7kHz and (ii) a 6.5kHz. Condition (i) is a loose configuration which corresponds precisely to the band constrain in Taori \textit{et al.} while condition (ii) is stricter.
 
\begin{figure}[h]
\centering
\includegraphics[width=0.7\linewidth]{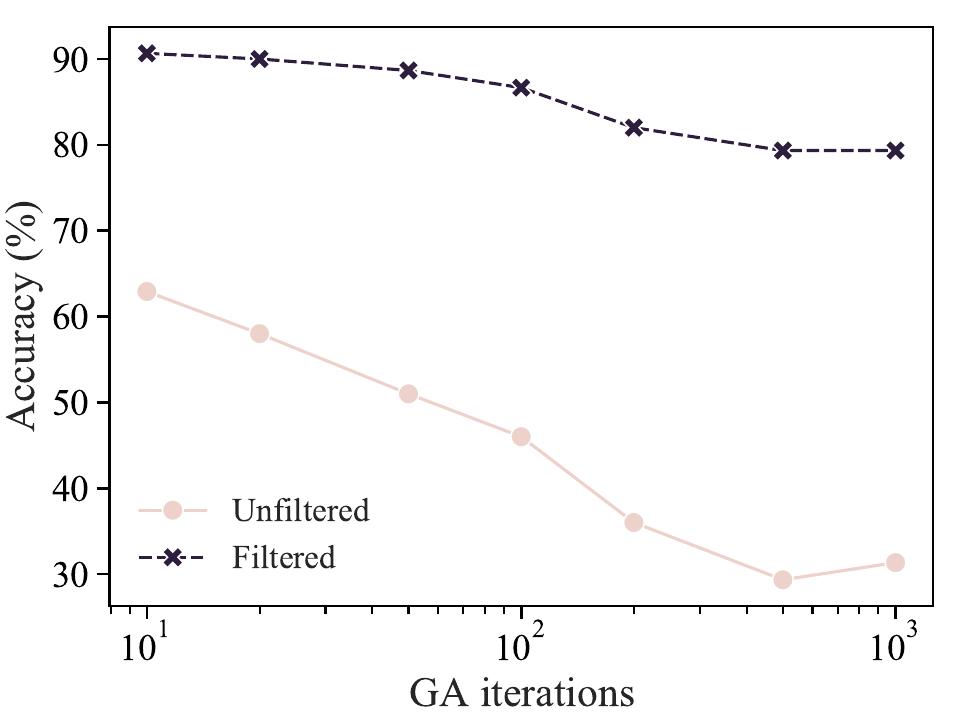}
\caption{\footnotesize \footnotesize \textit{Nyquist frequency enforcement:}
\textbf{i. Unfiltered} - Direct feeding of the adversarial audio, and 
\textbf{ii. Filtered} - Adversarial audio are passed through a 7kHz filter before processing.
}
\vspace{-10mm}
\label{fig:nyquist}
\end{figure}

Results for (i) are presented in Figure~\ref{fig:nyquist}. In the unfiltered case, the accuracy of the model quickly drops with the number of attack iterations. In the filtered case, the defender recovers most of the lost accuracy and the decrease slope is notably smaller. At 1000 iterations, the adversary is still capable of achieving a 15 percentage point accuracy drop. Still, the filtered accuracy is 79.3\%, a $2.5\times$ improvement over unfiltered accuracy at 31.3\%. For (ii), the accuracy on the filtered audio samples consistently remains above 90\% accuracy and drops only to 92\% for 1000 iterations.

\begin{tcolorbox}
\textbf{Take-away:} The generation of adversarial audio samples should not focus on imperceptibility (as argued in \S~\ref{subsubsec:overkill}), but instead on on evading pipeline-based detection schemes by satisfying pipeline imposed requirements for plausibility.
\end{tcolorbox}

\section{White-box Surreptitious attack}
\label{sec:wbox}

We showed how focusing on imperceptibility is not the appropriate goal for audio adversarial examples as it does not ensure that examples go undetected by humans (\S~\ref{subsubsec:overkill}) or algorithmically at inner pipeline stages (\S~\ref{subsec:constraints}). Thus, we define \textit{surreptitious} adversarial examples, and show how to craft them given \textit{white-box} access to the pipeline.

\subsection{Defining Surreptitiousness}
\label{subsec:covert}

The effectiveness of an adversarial example generation technique can be recast as its \textit{ability to subvert} the controls $\mathcal{C}$, \textit{i.e.}, to appear normal to the defender. These controls may be put in place at any stage of the pipeline, as we discussed in \S~\ref{subsec:constraints}. Let $\mathcal{M}$ be the subset of $\mathcal{C}$ corresponding to human plausibility and $\mathcal{F}$ the subset corresponding to pipeline-based detection. We taxonomize attacks on a scale inspired by physical penetration testing~\cite{ollam_key_decodingpdf_2019} which depends on $\mathcal{M}$ and $\mathcal{F}$. 

\begin{description}[noitemsep]
\item[Overt:] An attack is overt iff it makes no attempt at subverting the controls $\mathcal{M}$. An unbounded adversarial example, or simple high power frequency insertion (see Appendix~\ref{app:sec:frequency_insert}) falls under this category. An adversarial audio command that has not been obfuscated also falls within this category. 

\item[Covert:] An attack is covert iff it is undetectable by the controls in $\mathcal{M}$. However, this attack violates the controls $\mathcal{F}$. 
Existing imperceptible and obfuscation attacks fall under this category as they can be identified using the pipeline (\S~\ref{subsec:constraints}). A covert attack does not have to be imperceptible or fully obfuscated as long as it does not trigger the controls in $\mathcal{M}$. 

\item[Surreptitious:] A surreptitious attack is an attack that evades the entirety of the controls $\mathcal{C}$, \textit{i.e.}, does not leave conclusive evidence of attack.
\end{description}

In the rest of this work, we show that because of the pipeline, surreptitious audio attacks, \textit{i.e.}, attacks achieving typicality across all the stages, are possible but challenging.

\subsection{Strawman Surreptitious Attack}
\label{sec:strawman}
\label{subsec:diffbase}

In the vision setting, Tramer \textit{et al.}~\cite{tramer_adaptive_2020} use the feature level attack of Sabour \textit{et al.}~\cite{sabour_adversarial_2016} to bypass detection focusing on decision boundary anomalies of gradient-based attacks. Decision boundary anomalies are characterized by the different behavior of the logits of a benign and adversarial samples perturbed by gaussian noise.  
The feature level attack evades two constraints: (i) decision boundary anomalies, and (ii) human perception~\cite{sabour_adversarial_2016,tramer_adaptive_2020}, despite being defined on a single layer of the model. Hence, porting the attack to audio is a strong candidate for surreptitious adversarial examples: the adversary aims to optimize a similar objective with regard to a single stage of the pipeline.

\subsubsection{Methodology}
The Sabour \textit{et al.} feature matching attack~\cite{sabour_adversarial_2016} makes the internal representations of an input $x$ by the model layers $m_k$, $k\in[1,M]$, and the representations of the final layer's index $M$ close to that of a guide $g$, a sample whose features serve as the target (refer Equation~\ref{eq:feature_match}). To the best of our knowledge, we are the first to experimentally instantiate a gradient-based feature matching attack in audio. Note that the loss is computed on the $2$-norm but the perturbation is taken with the $\infty$-norm as in~\cite{sabour_adversarial_2016}. The adversary finds an adversarial audio $x'$ by solving for:
 \begin{equation}
 \label{eq:feature_match}
    \mathcal{L}(x,\delta)=\norm{m_k(x+\delta)-m_{k}(g))}_{2}
\end{equation}
\begin{equation}
    x' =\text{arg min}_{\norm{\delta}_{\infty}\leq \epsilon}(\mathcal{L}(x,\delta))
\end{equation}
Note that unlike internal representations of a computer vision model, a spectrogram is both human interpretable (\textit{i.e.,} humans are sensitive to frequency changes~\cite{schonherr_adversarial_2018}) and invertible~\cite{sturmel_signal_2011}. Making the spectrogram close to that of another sample will degrade the quality of a sample and, in the limit, the adversarial sample will correspond entirely with the guide if they have the exact same spectrum.

Implementing such an attack is nuanced. Audio samples are usually of different durations. The resulting spectrograms have variable time dimension, complicating the attack implementation. Contrary to computer vision, we thus need to (i) tackle the variable dimensions, and (ii) choose the stage at which the feature level objective is defined. For condition (i), we find suitable guides for each sample as follows: for each sample, we construct a candidate list with all the LibriSpeech~\cite{panayotov_librispeech_2015} samples of the exact same duration. We take a guide at random in the candidate list. This is possible because of the LibriSpeech test set construction. For the rare samples which do not have an exact duration match, we take the closest longer sample and truncate it. For condition (ii), we test the attack at 3 possible stages: at the spectrogram level, at earlier/shallower (model) internal features and for later/deeper (model) internal features. The shallow internal features are taken as the output of the first fully connected layer. The deep internal features are taken as the output of the $3^{rd}$ convolution layer. We discuss this choice of layers and its implications in our result analysis below. 

\subsubsection{Experiments} 
We implement a differentiable version of a spectrogram-based pipeline (\texttt{SBP}) in \texttt{Tensorflow 2.1} with its default pre-processing hyperparameters~\cite{nagrani_voxceleb_2017}. We use the entire input as this configuration provides the best benign accuracy as reported in~\cite{nagrani_voxceleb_2017}. We use the SincNet variant of the LibriSpeech dataset\footnote{\url{ https://github.com/mravanelli/SincNet/issues/25}}, with 2484 unique speaker labels. 

As a baseline method, we use the (untargeted variant of) FGSM, and 100 iterations of PGD to minimize the audio perturbations at stage 3~\cite{carlini_audio_2018}. Henceforth, we refer to this attack as the differentiable baseline attack. We measure the perturbation size \textit{both} in the digital space (stage 3) and spectrogram space (stage 1). To do so, we process the adversarial example through the pipeline and compute its ${\infty}$-norm distortion at both stage 1 and stage 3 of the processing. We perform each attack on a fixed subset of 1000 randomly selected audio samples. We use this same subset for all experiments in the paper. 

\begin{figure}[t]
 \centering
 \includegraphics[width=\linewidth]{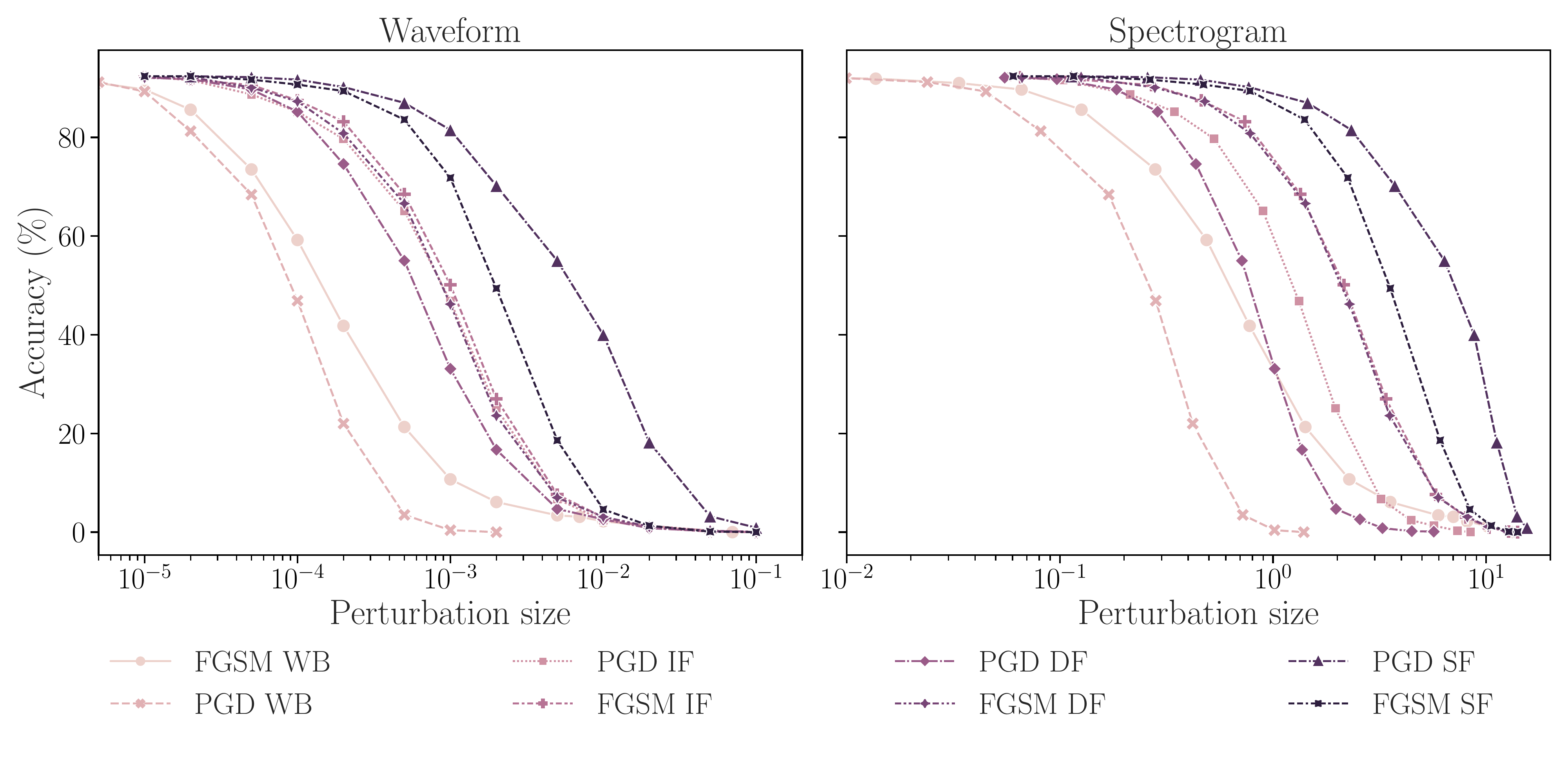}
 \caption{\footnotesize Comparison between the untargeted white-box differentiable pipeline (WB) and feature attacks for spectrogram features (SF), internal features (IF) and deep features (DF). The deeper the feature targeted, the smaller the necessary distortion is. Even in the best case, the distortion is significantly larger than in the baseline. 
 }
\label{fig:feature}
\vspace{-10mm} 
\end{figure}

\noindent{\bf Results:} Figure~\ref{fig:feature} reports the accuracy decrease against audio (left subfigure) and spectrogram (right subfigure) for equivalent perturbation sizes. As in Sabour \textit{et al.}~\cite{sabour_adversarial_2016}, the deeper the features, the more effective the attack is as features closest to the classification layer (stage 0) influence the outcome the most. For both model internal features and the spectrogram level features, we observe both attacks to be successful, although they both require different perturbation sizes. This suggests that feature extractors prepended to the model may be resilient to feature matching attacks since (i) the objective makes no attempt at reducing the spectral perturbation but only the audio perturbation with regard to the corresponding benign sample, and (ii) the objective leads to samples that have a spectral representation close to the guide (which may be very different from the benign sample). The attack is \textit{less surreptitious} than the differentiable pipeline baseline attack due to the resulting large spectral perturbation. 

\subsection{Joint Optimization}
\label{subsec:jo}

Given the negative results obtained by optimizing over a single pipeline stage, we propose a joint optimization between the traditional adversarial objective and a surreptitious objective. Inspired by the use of internal model representations in the feature matching attack, the surreptitious objective operates on internal representations of the pipeline. We aim to (i) find the minimal spectral detection threshold this attack can avoid, and (ii) understand the trade-offs that the adversary must take with this attack.

\subsubsection{Methodology}
In contrast to an attack objective designed to fool the model, the surreptitious objective conveys the need to avoid detection by minimizing perturbations on internal pipeline stages.

\paragraph{1. General surreptitious objective:} Ideally, the objective would be used on all possible pipeline internal features produced by the $p_i$. For an untargeted\footnote{In the white-box threat model, targeted and untargeted attacks can be achieved effectively. Hence, we restrict our evaluation to untargeted attacks. For the black-box case (\S~\ref{sec:snes}), we explore both options which behave very differently.} attack objective, for a benign audio sample $x$, the adversary maximizes the objective $\ell(m(x+\delta),y)$ with $\ell$ the standard cross-entropy. The general surreptitious objective is defined as a joint minimization on the norm of the perturbation induced over each independent pipeline stage and control (each with their own constraints), \textit{i,e.}, jointly over each independent $\norm{p_{i}(x)-p_{i}(x+\delta)}_{2}$ for $i\in[1,N]$, where $p_i$ denotes the $i^{th}$ of $N$ pipeline stages. 

\paragraph{2. Reduced surreptitious objective:} Note that there is a fundamental tension between the two components of the joint attack objective. One objective makes the perturbations adversarial, while the other limits their potency by keeping representations internal to the pipeline close to that of benign samples. Due to the lack of an easily manipulable closed form, this initial objective is difficult to implement and optimize as it requires to keep track of the perturbation and input for \textit{all} pipeline steps. Hence, we propose that the adversary reduces the scope of the surreptitious objective to major pipeline stage transitions, and resulting constraints, that will be observed by the defender as part of the controls $\mathcal{C}$. In the rest of this work, we restrict the surreptitious objective to the spectrogram\footnote{The attack objective is over audio while the surreptitiousness objective focuses on spectrograms resulting in optimization over two stages.}, a likely stage for control as shown in \S~\ref{subsec:constraints} because it is (i) in the frequency realm, (ii) the input of the model itself in \texttt{SBP}, and (iii) invertible contrary to model internal features. We further introduce a surreptitious factor $\lambda$ that allows the adversary to balance their covertness needs on the audio and spectrogram domains. We then experimentally explore the tensions in the objectives and effects of the variable $\lambda$. This yields the simplified joint optimization objective in Equation~\ref{eq:simplified_internal}, with $\sigma=p_1 \circ p_2 \dots p_{s}$ the function returning the power spectrogram for a sample $x$. 
 \begin{equation}
 \label{eq:simplified_internal}
    \mathcal{L}(x,\delta)=\underbrace{\lambda \ell(m(x+\delta),y)}_{attack} + \underbrace{(1-\lambda)(- \norm{\sigma(x)-\sigma(x+\delta))}_{2})}_{surreptitious}
\end{equation}

Observe that this objective by default only minimizes the spectral perturbation for a given audio perturbation. 
We thus additionally solve the optimization problem with regard to the digital audio (stage 3) as in Equation~\ref{eq:final_objective}:
\begin{equation}
    \max_{\norm{\delta}_{\infty}\leq \epsilon} \mathcal{L}(x,\delta)
\label{eq:final_objective}
\end{equation}

\subsubsection{Experiments} 
\label{subsubsec:joint_results}
\begin{figure}[t]
 \centering
 \includegraphics[width=\linewidth]{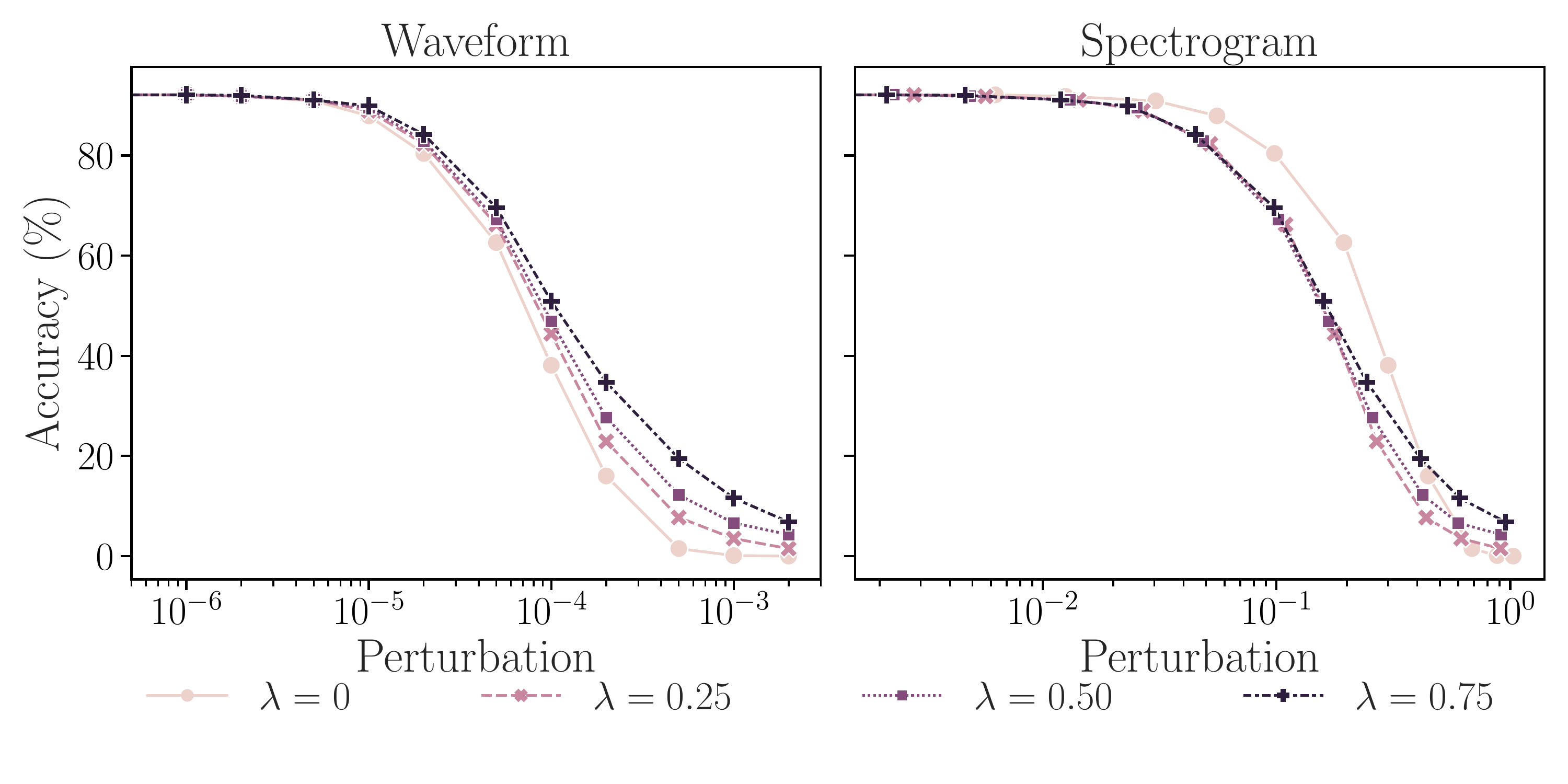}
 \caption{\footnotesize (Left) Audio distortion and (Right) spectrogram equivalent distortion (measured by the $\infty$-norm) comparison between the baseline ($\lambda=0$) and surreptitious attacks on \texttt{SBP}.
 }
 \label{fig:surrep}
 \vspace{-10mm} 
\end{figure}

We use the  \texttt{SBP} pipeline (previously described) and attack it using samples from PGD (ran for 100 iterations). For $\lambda=\{0,0.25,0.5,0.75\}$, we report the accuracy loss for given audio distortion and the equivalent spectrogram perturbation in Figure~\ref{fig:surrep}. The curve highlights that the audio perturbation needed for a given accuracy loss increases with $\lambda$. This is expected as the adversary objective must now additionally minimize the surreptitious objective.

The surreptitious objective significantly decreases the spectral perturbation induced. For an accuracy of 40\% this perturbation is decreased by a factor $2$. From reading the graphs, while the spectral perturbation is decreased by $2\times$, the audio perturbation increase is only $1.25\times$. The minimal spectral resolution threshold that the adversary can evade is therefore $2\times$ smaller than for the pipeline differentiation attack. 

\noindent{\bf Trade-Off:} For larger perturbation sizes, the curves cross with $\lambda=0$ eventually yielding the best possible attacks. With larger values of $\lambda$, the focus on the surreptitious objective forces larger audio distortions for a given accuracy loss, in turn producing larger spectral distortion. This means that the best configuration for the adversary is $\lambda=0.25$. We conclude that the regime of interest for this surreptitious attack requires audio perturbations to have a norm smaller than $0.001$, before the crossover happens. Indeed, as shown in the next section, for the LibriSpeech dataset, large audio perturbations with norms greater than $0.002$ would be covert, or they would be overt when the norm is larger than $0.1$.

\begin{tcolorbox}
\noindent{\bf Take-away:} A larger audio perturbation \textit{does not necessarily} translate to a larger spectral perturbation as the curves for $\lambda=0.25$ and the differentiable pipeline show. 
Thus, the adversary can trade spectral perturbation for audio perturbation making surreptitious attacks possible, as long as the resulting audio remains plausible. However, focusing solely on the surreptitious objective leads to a sub-optimal attack (as shown $\lambda=0.75$ which actually increases the spectral perturbation).
\end{tcolorbox}

\subsection{Human Evaluation}
\label{subsec:mturk}

Since our joint optimization introduces higher audio distortion than prior work, we evaluate the resulting audio samples with an Amazon Mechanical Turk (MTurk) study. Our goal is to verify that in addition to satisfying constraints from the pipeline, our attack samples remain plausible for trained humans. We obtained IRB approval prior to this experiment. We paid each participant \$2.5 for a task taking 25 minutes to complete on average, collected no participant private information, and ensured the volume of the audio samples was moderate to minimize the risk of any harm.

\noindent{\em MTurk for audio adversarial example evaluation:} The relevance of MTurk studies for \textit{imperceptibility} evaluation of adversarial examples is a debatable topic. This is due to the coarseness of imperceptibility definitions (\textbf{P1}); subjects' listening sensitivity and audio equipment quality are hard to control using MTurk~\cite{abdullah_sok_2020,abdullah_hear_2019}. Thus, imperceptibility claims should be asserted with stringent procedures including soundproof chamber with controlled headphones~\cite{schonherr_adversarial_2018}. Nonetheless, our study's aim is not to evaluate imperceptibility but rather the \textit{weaker requirement of plausibility}. We argue that the uncontrolled parameters are irrelevant for plausibility making an MTurk study adequate in our case.

\subsubsection{Methodology} 
We identify three limitations of the noisy and clean labels used (to label sound samples) in prior work~\cite{devil,qin_imperceptible_2019,chen2019real}. First, different types of noise exist, some of which like white noise, may naturally occur. Second, a sample can be non-plausible for a given application despite being clean: songs are suspicious inputs for a voice application (\S~\ref{subsubsec:overkill}). Third, benign audio dataset samples may in practice be quite noisy, making the assessment unfair in favor of the attack. Thus, we use the ratings \textit{(i) normal, (ii) abnormal, and (iii) unsure} which more closely align with the adversary's need of plausibility while simplifying participants' training as these ratings avoid exposure to our specific nomenclature. 

Unlike prior work~\cite{qin_imperceptible_2019,chen2019real,kreuk_fooling_2018}, to mimic trained experts and measure \textit{plausibility}, our study consists of a training and a testing phase. During the unskippable training phase, workers listen to 6 samples and are provided the expected rating of each sample. Two samples are benign, two are from overt attacks (refer Appendix~\ref{app:sec:frequency_insert}) and 2 from the differentiable baseline (refer \S~\ref{subsec:diffbase}). 

During the test phase, the participants must rate 30 samples. We choose the benign samples so that (i) two are inherently noisy, (ii) two are inherently clean, and (iii) the remaining (two) are randomly selected. In contrast, the samples used to generate the attacks are manually curated to ensure that they are not overly noisy before the attack is performed (as done in prior work). Unlike the attacks provided at training time, the test time attack samples are chosen across the spectrum of attacks from Abdullah \textit{et al.}’s SoK~\cite{abdullah_sok_2020}. They include (A1) the Devil whisper attack~\cite{devil} (an obfuscated attack), (A2) a differentiable pipeline attack (baseline)~\cite{carlini_audio_2018}, (A3) a feature attack~\cite{sabour_adversarial_2016}, (A4) an overt sine wave insertion attack (see Appendix~\ref{app:sec:frequency_insert}), and (A5) our joint optimization attack (\S~\ref{subsec:jo}). The order of the samples is random and is the same for all participants. 

For attacks (A2-4), we use $\epsilon=0.001$ and $\epsilon=0.1$ which correspond respectively to the maximum audio perturbation in the regime of interest for the joint optimization (\S~\ref{subsubsec:joint_results}) and evident tampering for over-the-air attacks in prior work~\cite{chen2019real}. For the feature attack (A3), a number of generated samples were not effectively adversarial due to the lower potency of the attack at $\epsilon=0.001$. Hence we had to manually curate the samples we retained. Observe that in practice, an adversary could further improve the attack by manually selecting the best sounding samples. 

To ensure that participants do not respond at random, we use three control samples consisting of a single non-voice frequency pulse generated with the \texttt{sox} utility~\cite{noauthor_sox_nodate} at 440, 250 and 800Hz. We additionally require participants to confirm that they were able to properly hear the samples and that they did not respond at random. We reject any participant that rates any control sample as normal, states that they responded randomly or stated that they could not hear the samples. Out of 150 submitted tasks each corresponding to independent human workers, 119 pass our stringent rules.

\subsubsection{Results}
\begin{figure}[t]
 \centering
 \includegraphics[width=\linewidth]{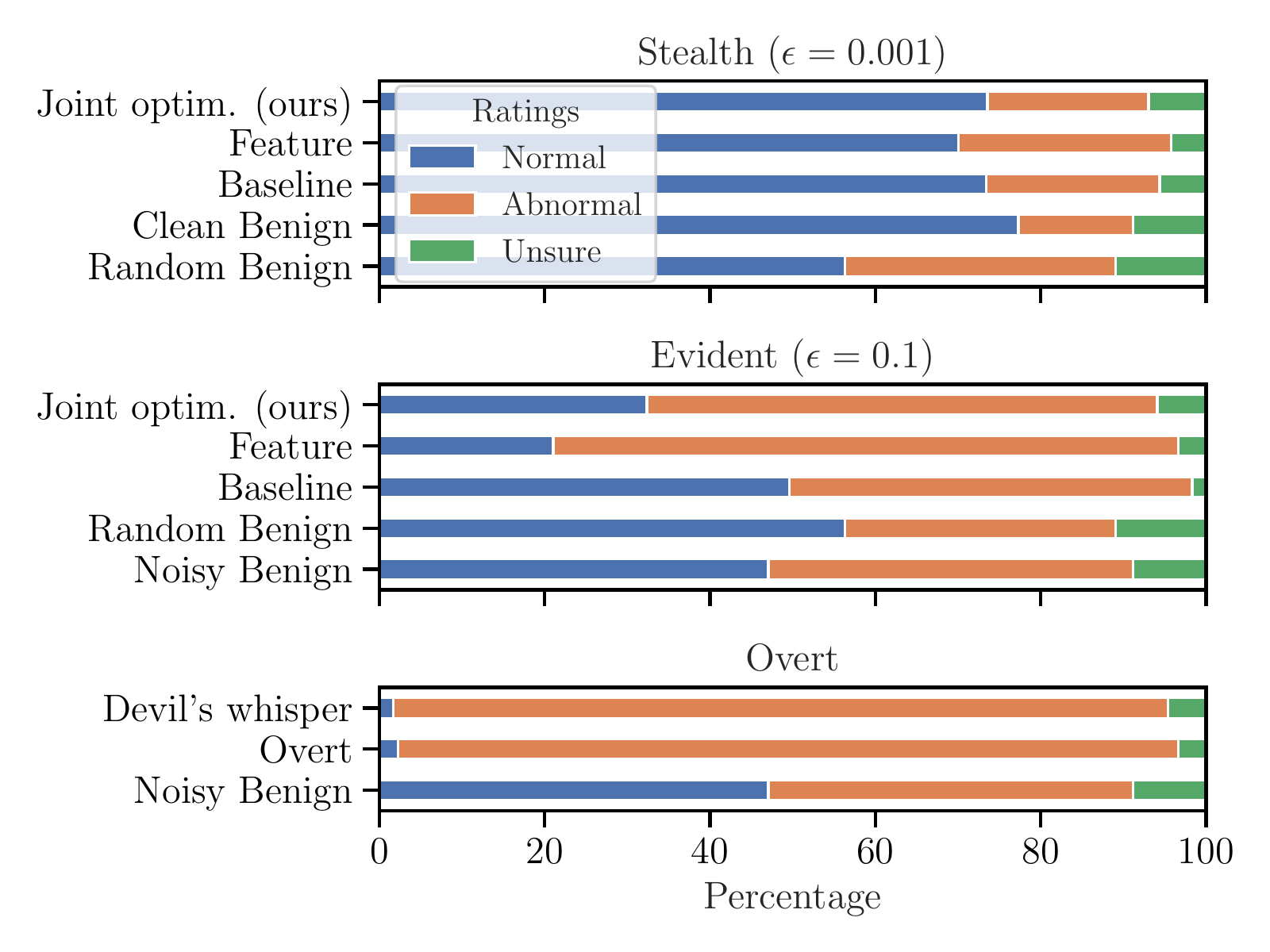}
 \caption{\footnotesize \textit{MTurk plausibility results.} The joint optimization attack has similar plausibility as previous attack techniques. }
 \label{fig:mturk}
  \vspace{-10mm}
\end{figure}

We plot the ratings breakdown per attack class and perturbation size in Figure~\ref{fig:mturk}. Lightly trained MTurkers were easily able to identify the overt frequency insertion and Devil's whisper attacks, rating them as abnormal in 94.4 and 93.7\% of cases against 46.22\% for noisy benign samples. We conclude that both attacks fall in the overt category as they are identified by more than 90\% of our human listeners. This is despite the frequency insertion attack (see Appendix~\ref{app:sec:frequency_insert}) being not intended to evade human perception and much cheaper to perform for an adversary than the Devil's whisper obfuscated attack. 

\noindent{\bf Take-away 1: Undetectable inputs should be within the distribution of plausible inputs.} Results in the overt setting confirm that an attack may target human perceptibility, yet are often identified as abnormal \textit{i.e.}, songs or noise cannot be used to mask an ASI attack. 

In the evident tampering configuration ($\epsilon=0.1$), all attacks performed were identified as abnormal significantly more often than the randomly selected benign samples which were only rated abnormal in 30.04\% of cases. The feature attack performed the worst and was rated abnormal in 75.63\% of responses. With a distortion $100\times$ larger than their intended use range, the joint optimization samples were recognized as abnormal in 61.77\% of cases, 15 percentage points below noisy benign samples. The pipeline differentiation attack was polarizing with only 1.68\% of responders unsure of the rating they should provide versus 9.24\% for noisy benign samples. Still, MTurkers rated the pipeline differentiation attack normal by 5 percentage points more than noisy benign samples.  

\noindent{\bf Take-away 2: Audible perturbations may produce plausible inputs.} The evident tampering configuration results confirm our perspective that (i) feature matching attacks behave very differently in the pipeline-heavy audio domain when compared to feature matching attacks in computer vision, (ii) plausibility can be achieved with clearly audible adversarial samples as shown by our baseline, and (iii) the joint optimization objective may prove detrimental to the adversary for larger perturbation sizes (recall that the baseline has lower spectral perturbation than the joint optimization attack for $\epsilon\geq0.001$ as the perturbation curves crossed over in Figure~\ref{fig:surrep}). Nonetheless, if the adversary uses these larger perturbation sizes they are not trying to remain stealthy or imperceptible to humans anyway.

In the stealth configuration ($\epsilon=0.001$), our joint optimization and the pipeline differentiation attack performed comparably on the normal ratings. Our joint optimization attack was rated abnormal by 1.46 percentage points less than the pipeline differentiation with most of the difference (1.33 percentage points) surfacing in the unsure category. Compared to benign samples, both attacks were rated normal within 4.75 percentage points from the clean (best) samples and at least 13.74 percentage points more than the random samples. The feature attack still performed the worst, trailing 6.23 percentage points behind our joint optimization attack in abnormal ratings but still significantly better than random benign samples. 

\noindent{\bf Take-away 3: Better use of perturbation budget is to avoid detection across all defender controls.} Last but not least, the stealth configuration results confirm that within the joint optimization attack's regime of interest, the adversary's perturbation budget would be better used to avoid detection across all defender controls rather than focusing solely on human perceptibility targets. Indeed, both the baseline and our joint optimization are rated normal almost as often as clean benign samples, despite higher audio perturbation for our attack than the baseline. Concurrently, our attack results in spectral perturbations twice as small as the baseline making it more surreptitious.
\section{Surreptitious Black-box Attack}
\label{sec:snes}

Next, we show that surreptitious attacks are \textit{challenging} in the black-box setting. We consider an adversary who does not have white-box knowledge of the pipeline and thus cannot explicitly model the pipeline constraints to directly attack the digital audio (refer \S~\ref{subsec:snes_intro}). Direct black-box attacks in the literature such as finite differences or NES~\cite{chen2019real} \textit{cannot} be used to mount surreptitious attacks with regard to the pipeline. These attacks approximate the white-box computation of gradients~\cite{qin_imperceptible_2019,carlini_audio_2018} with regard to inputs. Such methods cannot estimate intermediate features utilized in the surreptitious objective. To address surreptitiousness, we must therefore rethink black-box attacks through the lens of transferability. 

\subsection{SNES Formulation}
\label{subsec:snes_intro}

Recall that pipelines introduced in \S~\ref{sec:background} are diverse, both in terms of the pre-processing steps as well as the models used for classification. Thus, attacking such models effectively requires extensive white-box knowledge (such as access to internal features), especially for surreptitious attacks. At a high level, all models architecture are CNNs applied to the same task (TDNNs are a specific form of CNN) and are thus likely to learn functions representing similar input-output mappings. When white-box knowledge is not readily available (a realistic assumption in many practical settings), an adversary requires a surrogate. Note that \texttt{DBP}, the end-to-end DNN pipeline, embeds the traditional audio processing features such a Mel log scale used in other pipelines into the first layer of a CNN, and can approximate both \texttt{ABP} and \texttt{SBP}. Thus, black-box adversaries could use \texttt{DBP} as a differentiable replacement. The ability to do so implies that the pipelines are \textit{equivalent}, defined \textit{informally} as them performing the same function with limited component variation (\textit{e.g.}, they all use CNNs). If pipelines are equivalent, an adversary is more likely to be able to rely on the well known transferability property of adversarial examples~\cite{szegedy_intriguing_2013,papernot_practical_2016} \textit{i.e.,} adversarial examples crafted in one pipeline \textit{remain adversarial} for others.

To re-state our observations thus far: (i) white-box knowledge enables access to internal features, (ii) publicly available \texttt{DBP} is often a good approximation to properietary \texttt{ABP} and \texttt{SBP}, due to the notion of pipeline equivalence, and (iii) transferability of adversarial examples is likely facilitated by (ii). These points cummulatively lead to the design of the \textit{Simple, Natural End-to-End Surrogate (SNES) attack}. This allows an adversary to directly attack the audio input simply by backpropagating through the end-to-end differentiable surrogate. 

\subsection{Transferability Accross Pipelines}
\label{subsec:covert_snes}

We evaluate the efficacy of our approach by crafting adversarial examples on the \texttt{DBP} pipeline, and transferring them to the \texttt{SBP} and \texttt{ABP} pipelines. Details are presented below. 

\subsubsection{Methodology} 
We train \texttt{DBP} using the LibriSpeech dataset \cite{panayotov_librispeech_2015} and attack it using untargeted variants of both FGSM, and PGD (100 steps). Recall that \texttt{DBP} is implemented using SincNet. As a 1D CNN, SincNet operates on context windows of length 200 to 375\textit{ms}. During training these windows are randomized to regularize training. At test time, SincNet creates an ensemble of CNNs where each element of the ensemble operates on its own context window. By default the ensemble is created on context windows shifted by 10\textit{ms} resulting in high overlap. As part of our attack, we set the shift to zero and attack each context window independently using $\infty$-norm variants of FGSM and PGD. We concatenate all the attacked context windows. Such an $\infty$-norm attack is more granular than a single perturbation applied to the entire audio sample, and is consequently \textit{more effective}.

We use the same 1000 random samples as before (\S~\ref{sec:wbox}). The generated adversarial samples are saved as 16kHz, 16 bit LPCM \texttt{wav} files and fed to \texttt{SBP} and \texttt{ABP}. This is the least restrictive encoding for our adversary since the linear grid provides the adversary with the most granular control on the size and locations of the perturbations introduced. Our \texttt{DBP} implementation is a fork of the open source \texttt{PyTorch} implementation.\footnote{\url{https://github.com/mravanelli/SincNet}} We do not modify the training procedure and only add the attack implementation. \texttt{ABP} is implemented in \texttt{TensorFlow 2.1} atop the differentiable spectrogram computation in \texttt{SBP} with a differentiable MFCC implementation based on~\cite{carlini_audio_2018}.

\begin{figure}[t]
 \centering
 \includegraphics[width=0.7\linewidth]{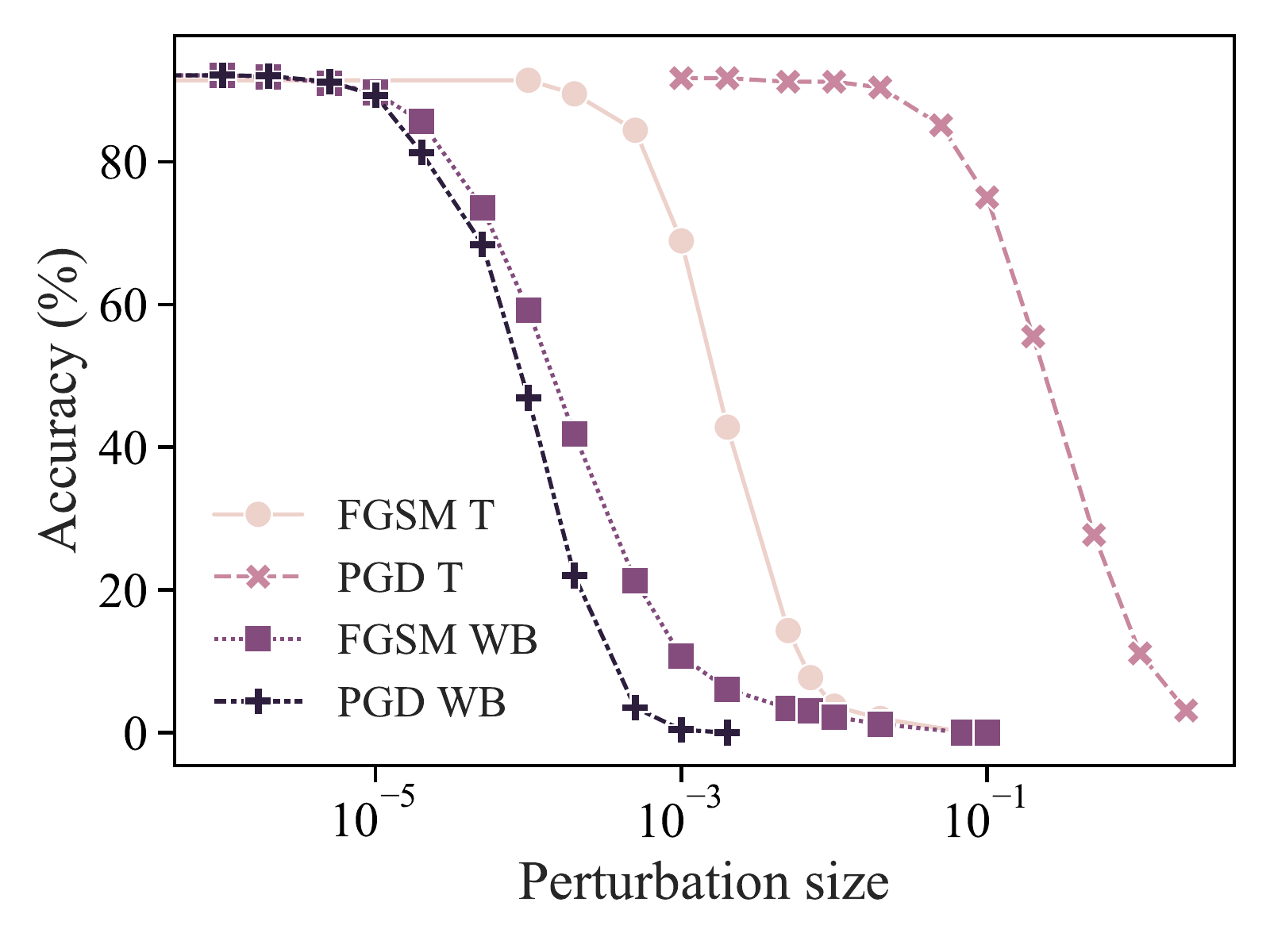}
 \caption{\footnotesize \textit{Untargeted attacks on waveforms.} Audio FGSM distortion (measured by the $\infty$-norm) comparison between the differentiable pipeline and transferability-based SNES attack. Comparison between FGSM and PGD attacks for the white-box (WB) direct attack on Vggvox (\texttt{SBP}) vs the SNES attack from SincNet (\texttt{DBP}) to Vggvox (denoted as T for \textit{Transfer}). 
 }
 \label{fig:perturbation_audio}
 \vspace{-10mm} 
\end{figure}

\begin{figure}[t]
 \centering
 \includegraphics[width=0.7\linewidth]{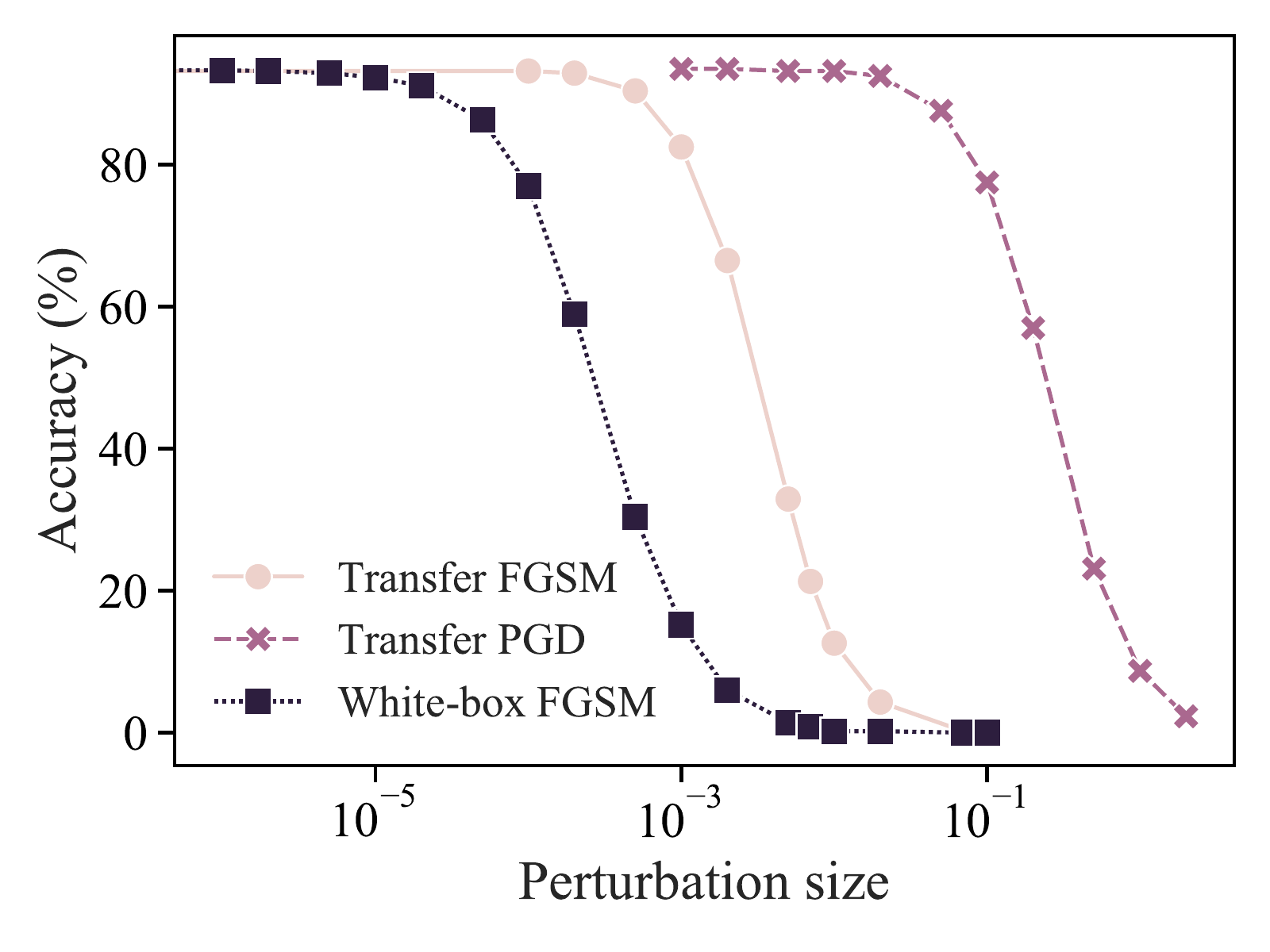}
 \caption{\footnotesize \textit{Untargeted attacks on waveforms.} Comparison between the white-box direct attack on X-vector (\texttt{ABP}) and the SNES attack from SincNet (\texttt{DBP}).}
 \label{fig:perturbation_audio_x_vec}
 \vspace{-10mm} 
 \end{figure}

\subsubsection{Results} 
\begin{figure}[t]
\centering
\includegraphics[width=0.7\linewidth]{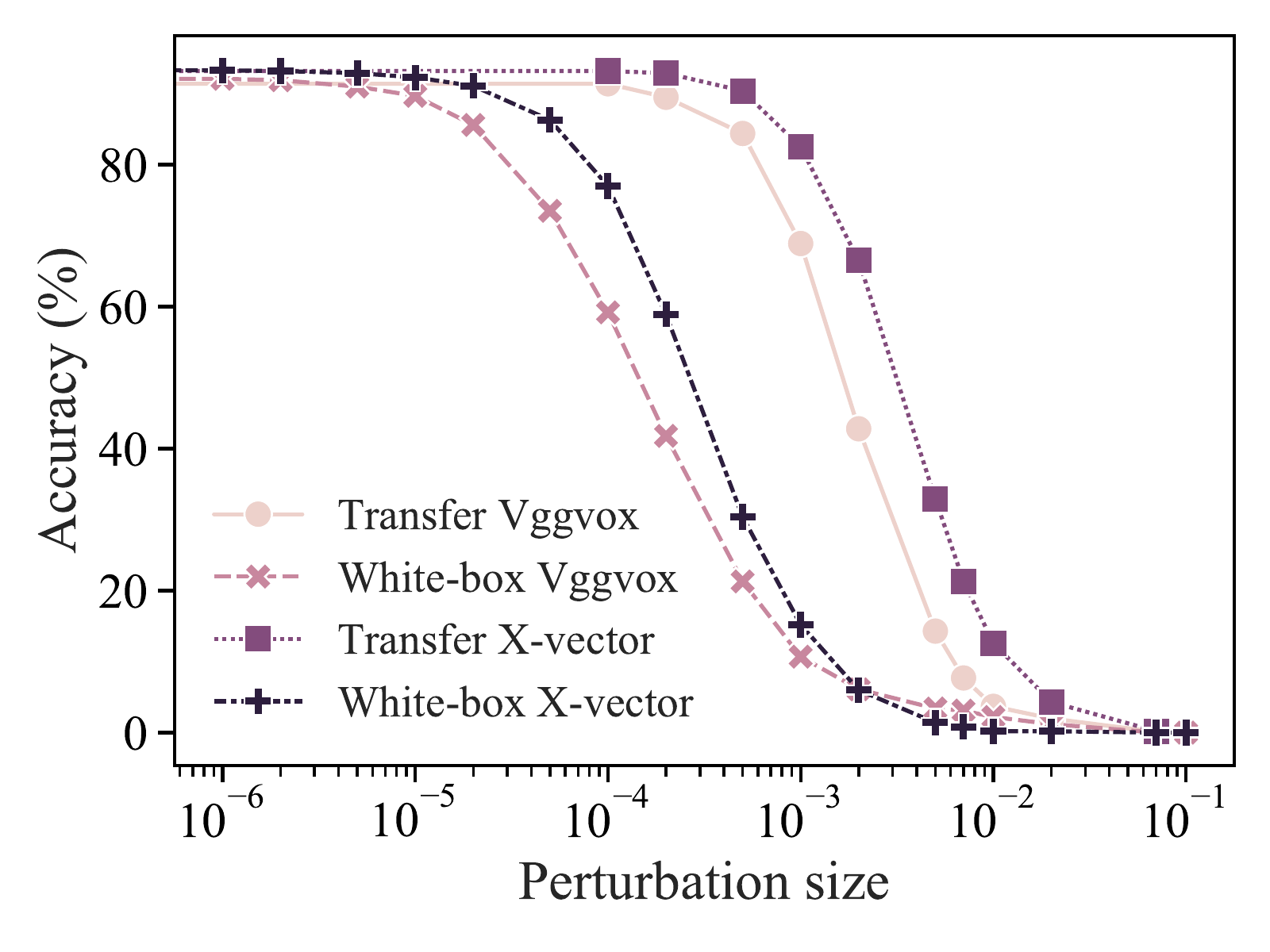}
\caption{\footnotesize \textit{Vggvox (\texttt{SBP}) vs X-Vector (\texttt{ABP}).} Comparison between Vggvox and X-Vector pipelines using the FGSM attack for the white-box direct attack and transferability-based SNES attack from SincNet (\texttt{DBP}) to either Vggvox or X-Vector. 
}
\label{fig:vggvox_vs_xvector}
\end{figure}

Experimental results for the SNES attack on \texttt{SBP} are presented in Figure~\ref{fig:perturbation_audio}. These results show that untargeted adversarial examples successfully transfer between pipelines. For example, the FGSM variant of the SNES attack can reduce the model accuracy from 93\% to 0\%, validating our claim of pipeline equivalence. The white-box PGD attack is more effective than FGSM for a given perturbation size meaning that the adversary can trade-off computational time for less detectable adversarial examples. We proceed to compare the transferability of the FGSM variant of the SNES attack between \texttt{ABP} and \texttt{SBP}. The results are highlighted in Figure~\ref{fig:vggvox_vs_xvector} and suggest that \texttt{SBP} is easier to attack than \texttt{ABP} as smaller perturbation sizes consistently lead to higher accuracy degradation as shown by the curve relative positions. Explaining the higher transferability between certain pairs of pipeline is difficult, as shown in prior work on computer vision which proposed varying explanations for transferability~\cite{DBLP:journals/corr/TramerPGBM17, liu_delving_2017}. 

\begin{figure}[t]
\centering
\includegraphics[width=0.7\linewidth]{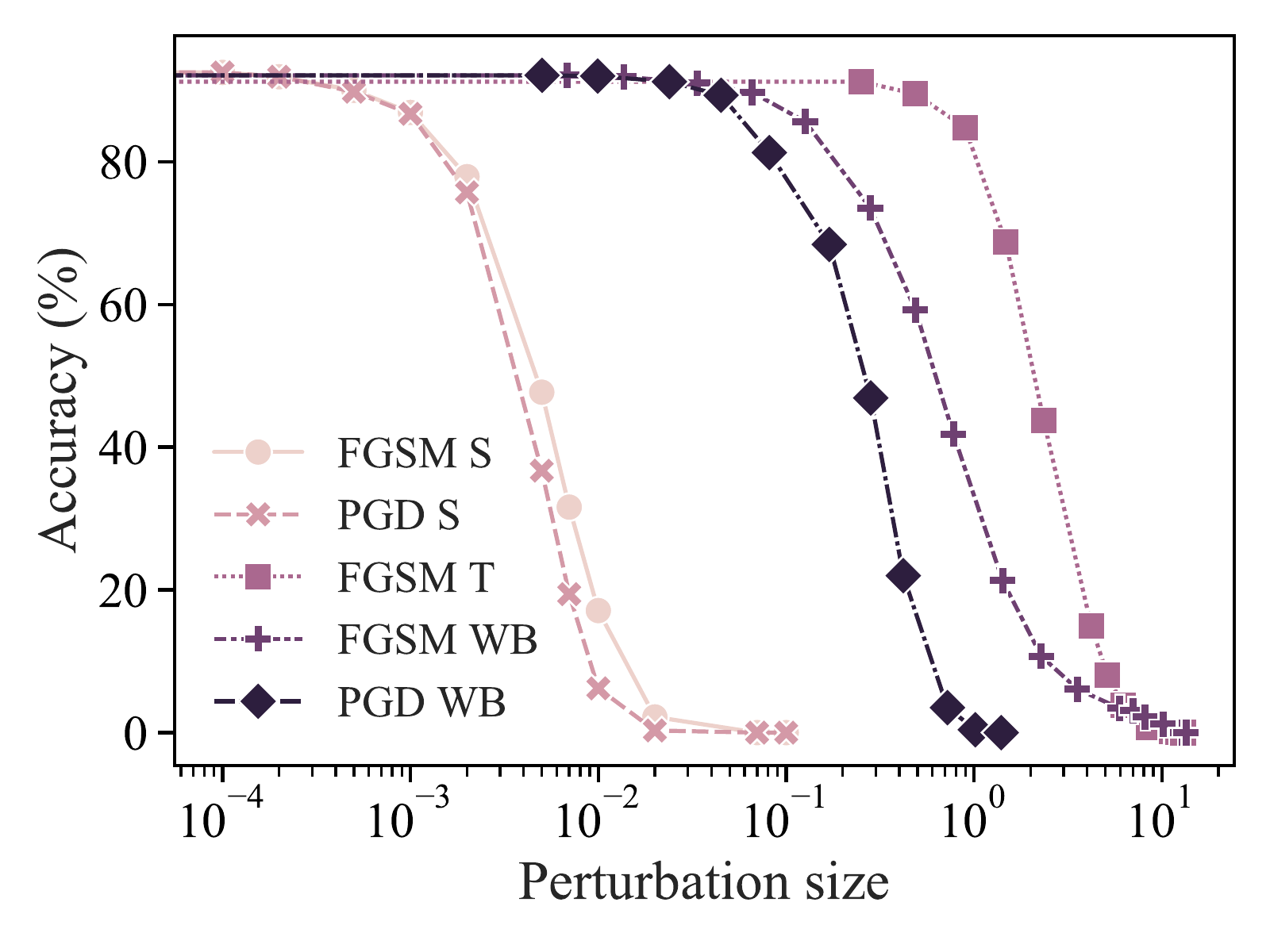}
\caption{\footnotesize \textit{Untargeted attacks in the spectogram stage.} Comparison of the accuracy vs perturbation size incurred by FGSM and PGD (100 iterations): (1) for the stage 1 attack on spectograms (S), (2) SNES attack (T), and (3) white-box (WB) baseline attacks on waveforms. The distortions are measured on the spectogram. The target pipeline is Vggvox.
}
\label{fig:perturbation_spectro_transfer}
\vspace{-10mm} 
\end{figure}

\noindent{\bf Trade-Off:} When we compare the accuracy of the target model for a given maximum $\infty$-norm perturbation on the audio sample incurred by the attack at the digital stage, the picture is quite different (Figure~\ref{fig:perturbation_spectro_transfer}). The perturbation incurred by the best 
SNES attack is one to two orders of magnitudes larger than the penalty incurred in the white-box attacks. In contrast to the whitebox case, transferability of the SNES PGD attack is worse than that of the SNES FGSM since with each iteration of the attack increasingly exploits the surrogate pipeline; this creates an adversarial example more specific to the surrogate and less generalizable to the target pipeline. We observe a similar pattern for attacks on \texttt{ABP} in Figure~\ref{fig:perturbation_audio_x_vec}. 

\begin{tcolorbox}
\noindent{\bf Take-away:} With SNES, the adversary trades off simplicity for a perturbation budget by an order of magnitude larger than in the white-box case. The adversary exchanges attack complexity for effectiveness and covertness. Moreover, by using PGD with SNES, the adversary can no longer trade-off computational time for less detectable adversarial examples. Since the audio perturbation is larger than the baseline, and similar to the feature matching attack, we say that the SNES
attack is \textit{covert}. Since the spectrogram perturbation is an order of magnitude larger than the baseline, the default SNES is certainly not \textit{surreptitious} for \texttt{ABP} and \texttt{SBP}.
\end{tcolorbox}

\subsection{Intermediate Feature Alignment}
\label{subsec:jo_snes}

We now explore whether enhancing SNES with a joint optimization objective on the internal layers of \texttt{DBP} results in a surreptitious attack. We show that this modification actually \textit{decreases} the attack's surreptitious potential. 

\subsubsection{Methodology} 
\begin{figure}[t]
 \centering
 \includegraphics[width=\linewidth]{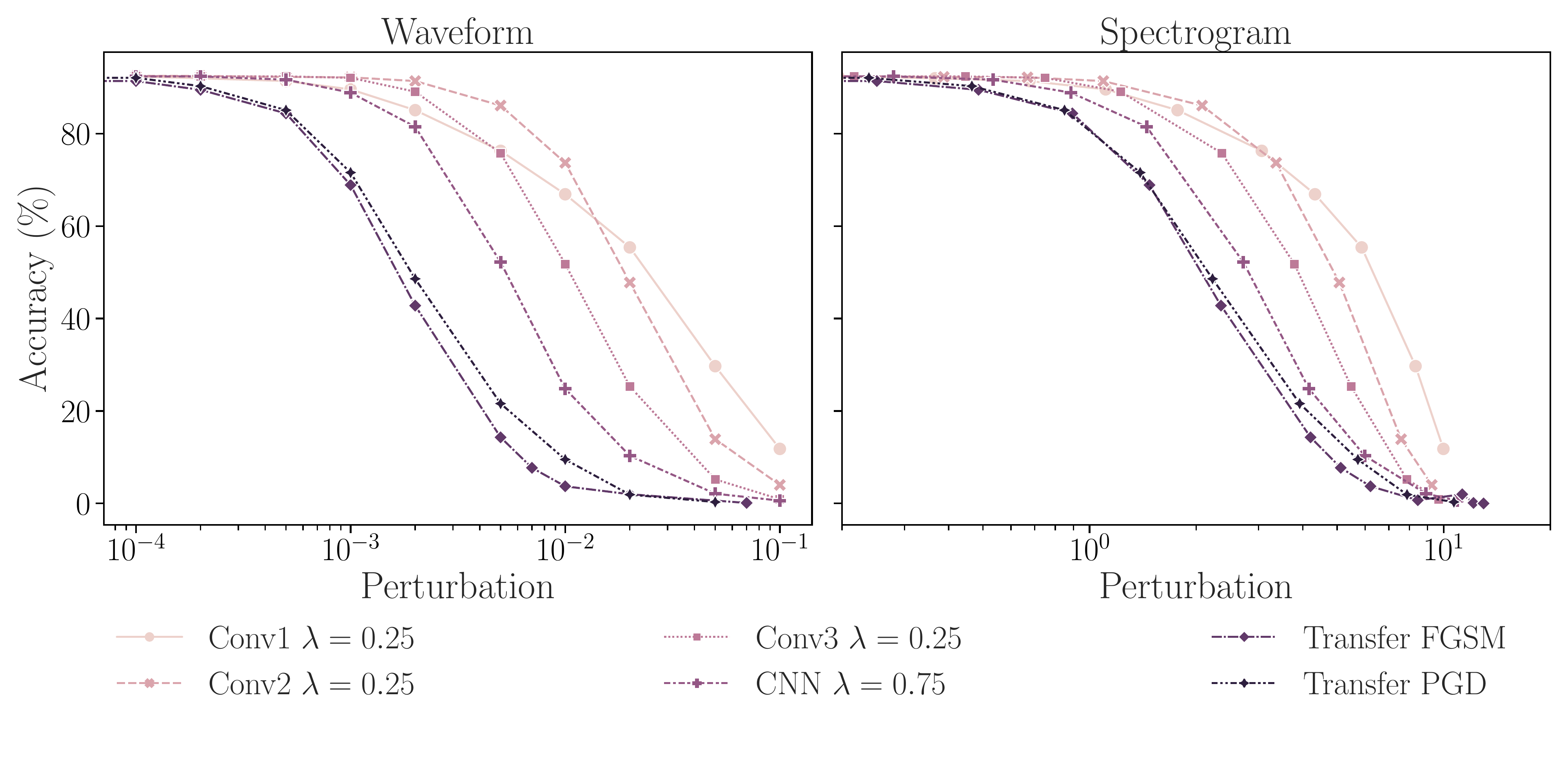}
 \caption{\footnotesize (Left) Audio distortion and (Right) spectrogram equivalent distortion (measured by the $\infty$-norm) comparison between the baseline ($\lambda=0$) and SNES joint optimization attacks on \texttt{SBP}.
 }
 \label{fig:jo_snes}
\vspace{-10mm} 
\end{figure}

The formulation of the objectives is the same as in the white-box case (\S~\ref{subsec:jo}). The only difference is that the surreptitious objective is performed on the output of internal layers of \texttt{DBP} rather than on an explicit spectrogram. \texttt{DBP} can conceptually be divided as a CNN followed by two fully connected DNNs. We test our attack on the output of each of the 3 convolution layers of \texttt{DBP} as well as the output of the CNN component. We test these variants of the SNES attack with \texttt{SBP} as our target and $\lambda=\{0.25,0.5,0.75\}$. To avoid cluttering Figure~\ref{fig:jo_snes}, we only report the best attack configurations for each internal layer.

\subsubsection{Results}
On \texttt{SBP} the additional surreptitious objective results in larger spectral perturbation sizes. The best strategy for the adversary is thus \textit{not to have} the surreptitious component in the attack objective. As in the white-box case, using $\lambda=0.25$ is the best configuration for all but one internal layer. Note, however, that the white-box joint optimization attack does increase surreptitious potential for well chosen $\lambda$ (\S~\ref{subsec:jo}) and the joint optimization is successful to reduce perturbation norms on internal layers of \texttt{DBP} (not reported). This leads us to attribute the failure of the SNES joint optimization attacks to the transfer step because, contrary to the high level pipeline, intermediate features of both \texttt{DBP} and \texttt{SBP} are not equivalent in the models we tested. 

\begin{tcolorbox}
\noindent{\bf Take-away} Contrary to model-level transfer attacks, black-box \textit{pipeline-level} surreptitious attacks have the strong requirement that intermediate features of both \texttt{DBP} and \texttt{SBP} be close. Under distillation alone as we experimentally tested, this requirement is not satisfied.
\end{tcolorbox}

\section{Related Work}

\noindent{\bf Audio pipelines for speech recognition.} 
Traditional human voice recognition systems used hand engineered features such as Mel Frequency coefficients (MFCC)~\cite{davis_comparison_nodate} with Voice Activation Detection (VAD) followed by a machine learning model, initially a Gaussian Mixture Model (GMM). Over recent years, the traditional models have been replaced with Deep Neural Networks ~\cite{snyder_deep_2017,snyder_x-vectors_2018}. Recent approaches have focused on reducing the need for hand-engineered features by restricting the extracted features to simple power spectrograms followed by Convolutional Neural Networks (CNNs)~\cite{nagrani_voxceleb_2017,Nagrani20}. Lastly, end-to-end deep neural network architectures for voice processing have surfaced ~\cite{tuske_acoustic_nodate,palaz_analysis_nodate,ravanelli_speaker_2019}. Sturmel \textit{et al.}~\cite{sturmel_signal_2011} discuss the precise conditions that govern the unique reconstruction of a signal from its STFT using the Griffin-Lim algorithm and its improvements.
 
  \noindent{\bf Adversarial examples \& audio attacks.} For white-box attacks against audio pipelines, we need to differentiate through the pipeline as in the work of Carlini \textit{et al.}~\cite{carlini_audio_2018}. In the image domain, Tramer \textit{et al.}~\cite{tramer_adaptive_2020} propose a modified version of the feature attack of prior work~\cite{sabour_adversarial_2016} which does not exhibit common artifacts of the FGSM and PGD attacks. 
 
 The untargeted transferability of adversarial examples has been widely explored in computer vision~\cite{papernot_practical_2016,szegedy_intriguing_2013,liu_delving_2017}. Targeted transferability is a known hard problem that has been less explored~\cite{sharma_effectiveness_2019, dong_boosting_2018,liu_delving_2017}. The transferability of adversarial examples in the audio domain is a relatively unexplored area with few contributions~\cite{kreuk_fooling_2018,abdullah_sok_2020}.
 
Abdullah \textit{et al.}~\cite{abdullah_sok_2020} provide an SoK on adversarial machine learning for MFCC ASR pipelines. Attacks on audio systems can be divided in two classes. The first class is composed of model centric attacks~\cite{carlini_hidden_2016,carlini_audio_2018,taori_targeted_2019,qin_imperceptible_2019,marras_adversarial_2019,kreuk_fooling_2018,vaidya_cocaine_nodate}. The other category is composed of signal centric attacks~\cite{abdullah_hear_2019,abdullah_practical_2019,abdullah_sok_2020,DBLP:conf/ccs/LiW00020}. ASR represents the vast majority of these with only few specifically considering the ASV setting of voice biometrics~\cite{kreuk_fooling_2018,marras_adversarial_2019}. Direct black-box attacks have been presented and are based on genetic algorithm, finite differences~\cite{taori_targeted_2019} or NES~\cite{chen2019real}.

\noindent{\bf System perspective.} A number of recent works have considered the security of ML components in their wider system context. At a high level, Evtimov \textit{et al.}~\cite{evtimov_security_2020} analyze the limitations of considering ML models in isolation and encourage system security views of the ML vulnerabilities. Prior work~\cite{gilmer_motivating_2018, pierazzi_intriguing_nodate, 236320} consider hard pipeline constraints to improve adversarial example attacks on malware detectors. From the defender's perspective, by analyzing high level systems goals, Chandrasekaran \textit{et al.}~\cite{chandrasekaran_rearchitecting_2019} present hierarchical classifiers that can be used to provide gracious degradation of a model in adversarial settings. Our observation that a proactive defender can observe each step of the pipeline is similar in idea to statistical testing at each of the layers of a computer vision model~\cite{dknn, raghuram_detecting_2020,DBLP:conf/ndss/MaLTL019}. Sheatsley \textit{et al.}~\cite{sheatsley2021robustness} make similar observations, and propose a data-driven approach to learn constraints from data enabling robustness. Concurrent work by Hussain \textit{et al.}~\cite{waveguard} consider ASR systems. Their solution does not consider pipeline induced constraints and instead relies on the vulnerability of adversarial inputs to specific audio transformations (similar to randomized smoothing~\cite{yang2020randomized} for vision). Additionally, the adaptive attack in their work focuses on ensuring indistinguishability, which they fail to validate through an empirical human study.

\section{Recommendations}

Our work outlines limitations in threat models of prior adversarial ML (AML) research. This paints a pessimistic view of the risks of adversarial examples in ML. To improve future attack evaluation, we make the following recommendations:
\begin{enumerate}
    \item Align adversary goals and capabilities at the system level, not at the model level. Considering ML models in isolated fashion provides limited insight in risk implications of adversarial examples~\cite{dreossi_semantic_2018,evtimov_security_2020,gilmer_motivating_2018}. Deployment schemes should be taken into account as intrinsic component vulnerabilities may be exacerbated or mitigated by interactions within the system (2009 ISO standard on the security evaluation of biometrics~\cite{1400-1700_isoiec_2020-1}).
    \item Develop attack objectives based on adversary needs. Objectives should be \textit{no more and no less} than strictly necessary to achieve the system level goal. A particular example has been discussed extensively in prior work: attacking the analog audio signal (stage 4) is referred to as an \textit{over-the-air} attack. This threat model requires the adversary be \textit{geographically co-located} with the victim acquisition system which greatly limits the corresponding risk. This geographic constraint is in and of itself costlier than alternatives (stage 3) offered to an adversary, like client-server or API attacks.
    \item Do not rely on weak or artificially-limited defenders for a stealthy attack to be effective. Evading one detection method while making detection simple by another method does not achieve much for the adversary, especially if the evasion technique is expensive. A shift in perspective is needed to best envision stealthy attacks. The stealth party needs to protect the \textit{confidentiality of their attack} from the detector no matter the detection method, \textit{e.g.}, can not limit themselves to evade constrained humans. It is well known that defense evaluation should not rely on weaker attackers and should be evaluated against adaptive attackers~\cite{carlini2019evaluating}. Similarly and in a somewhat unusual way, stealthy attacks should be evaluated against adaptive detectors.
\end{enumerate}

\section{Conclusions}

We took an end-to-end perspective on ML pipelines for voice biometrics authentication (using ASI). 
We considered the case of a proactive defender that can perform analysis of the major stages of an audio pipeline. 
The focus on imperceptibility of the perturbations in prior work is a red herring as this opens the door for pipeline-based identification of the attacks. We thus introduce the concept of surreptitious attacks that defeat both humans and pipeline-based controls. We realize that the adversary only requires their samples to be plausible to human subjects, a weaker requirement than imperceptibility. In the white-box case, our joint optimization attack can instrument this weaker requirement to trade larger audio perturbation for smaller pipeline perturbations, while at the same time remaining plausible as demonstrated by our user study. Finally, we show that black-box surreptitious attacks are challenging as our surreptitious optimization objective requires access to internal features of the pipeline.

\section*{Acknowledgments}
We would like to thank the reviewers for their insightful feedback. We would like to thank members of the CleverHans lab in particular Ilia Shumailov, Lucas Bourtoule, Nick Jia and Ali Shahin Shamsabadi for useful feedback and discussion. This work was supported by CIFAR (through a Canada CIFAR AI Chair), by NSERC (under the Discovery Program and COHESA strategic research network), by DARPA through the GARD program, and by a gift from Microsoft. We also thank the Vector Institute’s sponsors. Varun Chandrasekaran was supported by the Lawrence H. Landweber fellowship.


\newpage
{\footnotesize 
\bibliographystyle{acm}
\bibliography{biblio}
}
\newpage
\appendix

\newpage 
\section*{Appendix}

\section{Common Preprocessing}
\label{app:sec:preproc}

\subsection{Analog to Digital Conversion}
\label{app:subsec:A2D}
An analog to digital converter contains at a high level at least two elements.
First is a low-pass anti-aliasing filter. Aliasing is a spectrum folding phenomenon occurring in digital signals for insufficient sampling frequencies. 
This filter prevents aliasing by eliminating the signal frequencies above the Nyquist frequency responsible for the spectrum folding.

\begin{figure}[H]
\centering
\includegraphics[scale=0.33]{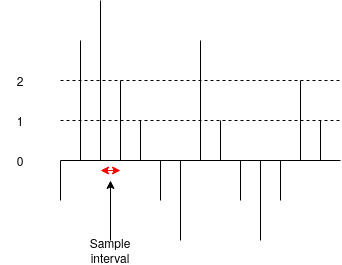}
\caption{\small Quantized signal. The values taken are integers, the space between two values defines the sample rate.}
\label{app:fig:quatization}
\vspace{-5mm}
\end{figure}

Second is a discretization component. A continuous analog signals cannot be represented with infinite precision. Thus, the signal is discretized, as illustrated in Figure~\ref{app:fig:quatization}, with respect to time, known as \textit{sampling}, and amplitude, known as \textit{quantization}, through a range of encoding schemes.

\paragraph{Quantization \& Encoding.}  The simplest encoding scheme is Pulse Code Modulation (PCM) which maps the waveform to an integer grid. For linearly spaced grids, this encoding is known as Linear PCM (LPCM). To maximize dictionary symbol usage, it is common in GSM codecs~\cite{noauthor_g7111_2020} to use non-uniform grid PCM like the $\mu$-law. In this case the grid is finer for symbols of interest to human intelligibility, and coarser outside. Voice applications are peculiar as it is possible to represent voice with high fidelity on constrained channels using either statistical correlations (\textit{e.g.}, ADPCM) in the signal~\cite{noauthor_g726_2020} or generative predictive models~\cite{lpc} encoding voice characteristics. Only coefficients of the predictive filter are sent over the network and the signal is reconstructed at reception. 

\subsubsection{Audio Codecs}
\label{app:subsubsec:codecs}
For most voice applications, sampling is done at 8kHz (narrowband) or 16kHz(broadband). 

\paragraph{Narrow band codecs}
The PSTN G.711 codec uses a PCM encoding using either A-law or $\mu$-law. The G.726 codec defines uses of ADPCM in narrow band ~\cite{noauthor_g726_2020}. The G.728 and G.729 standards use the 300Hz to 3400Hz range.

\paragraph{Broadband codecs}
The broadband extension of G.711 enables frequencies up to 7kHz ~\cite{noauthor_g7111_2020}. G.722 is defined by ITU-T to cover the 50Hz to 7kHz frequency band using ADPCM ~\cite{noauthor_g722_2020}. The G.722.2 codec also covers the 50Hz to 7kHz.

\subsection{Digital sound processing}
\label{app:subsec:digital_processing}
\begin{figure}[H]
\centering
\includegraphics[scale=0.33]{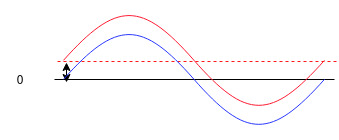}
\caption{\small DC offset removal on a sine wave. The DC offset shifts the mean of the signal to the dashed line.}
\label{app:fig:dc}
\end{figure}

\paragraph{DC Filter}
The dc filter normalizes the input audio by removing the mean amplitude of the waveform resulting in a signal with offset 0 (see Figure~\ref{app:fig:dc}). 

\begin{figure}[H]
\centering
\includegraphics[scale=0.33]{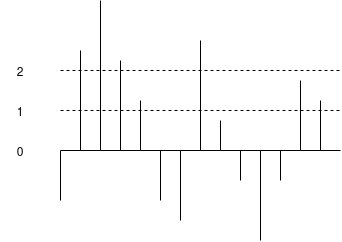}
\caption{\small Post dither insertion.}
\label{app:fig:dither}
\end{figure}

\paragraph{Dither}
Dither (see Figure~\ref{app:fig:dither}) prevents avoid a \textit{band effect} on the spectrogram due to quantization by inserting a small amount of noise to decorelate the quantization error.

\paragraph{Pre-emphasis}
\label{app:susubbsec:preemp}
Pre-emphasis filtering boosts the signal to noise ratio. Low frequencies are more likely to be high power than high frequencies. Hence, a high pass filter is applied to balance the relative power of the spectrum components.

\subsection{Spectral processing: STFT}
\label{app:subsec:Fourier}
\label{app:subsubsec:spectro}

\begin{figure}[t]
\centering
\includegraphics[scale=0.33]{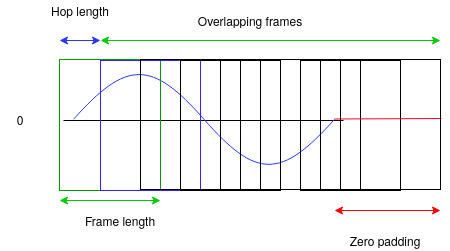}
\caption{\small The frame splitting operation. }
\label{app:fig:frames}
\end{figure}

\begin{figure}[t]
\centering
\includegraphics[scale=0.33]{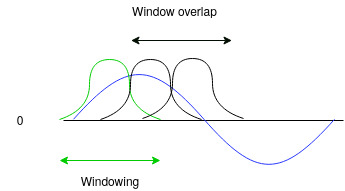}
\caption{\small The windowing operation.}
\label{app:fig:windowing}
\end{figure}

\paragraph{Frame splitting} To compute the Short Term Fourier Transform, the input signal is divided in frames (see Figure~\ref{app:fig:frames}). Due to the finite time horizon of the frame, this operation introduces unwanted frequencies which are limited by windowing (see Figure~\ref{app:fig:windowing}). VGGVox uses the Hamming window that strikes a balance between dynamic range and resolution.

\paragraph{Frame overlap} To compensate for the amplitude loss of windowing and provide better reconstruction, frames are overlapped. Vggvox uses a 25ms window with 10ms step and a 60\% overlap. The spectrogram produced is then normalized in mean and variance for each frequency bin.

\section{Model architectures}
\label{app:sec:architectures}
\paragraph{ABP: X-Vector and TDNN Architecture}

The X-vector model is a TDNN architecture~\cite{snyder_x-vectors_2018} operating on the frames of an MFCC input. Much like a 1-D CNN, in a TDNN the initial layers operate on context windows of the input with higher layers operating on larger contexts~\cite{waibel:phonemetr,DBLP:journals/neco/Waibel89}. We train this model from scratch on Librispeech, all parameters and hyperparameters are kept the same as in the X-vector paper~\cite{snyder_x-vectors_2018}. 

\paragraph{SBP: Vggvox}
Vggvox is a VGG-M CNN architecture modified to operate on spectrograms~\cite{nagrani_voxceleb_2017}. VggVox is implemeted as a fully convolutional network, \textit{i.e.}, the fully connected last three layers are implemented using equivalent convolution layers. We train our own version of Vggvox using using the Adam optimizer~\cite{DBLP:journals/corr/KingmaB14} and the provided speaker verification weights as a warm start\footnote{\url{https://github.com/a-nagrani/VGGVox}}. The batch size is 256 for training.

\paragraph{DBP: SincNet}
SincNet replaces the first layer of time domain 1D convolution filters in a otherwise standard CNN with \textit{Sinus cardinal} ($sinc$) functions and Mel log scale mapping. Since the Fourier Transform of a $sinc$ is a door, the network learns precise high-low cutoff frequencies for its filters~\cite{ravanelli_speaker_2019}. All training parameters are the same as in~\cite{ravanelli_speaker_2019}.\footnote{\url{https://github.com/mravanelli/SincNet}} 

\section{Hermitian Symmetry: Adaptive Adversaries}
\label{app:sec:adaptive_symmetry}

\begin{figure}[t]
\centering
\includegraphics[width=0.6\linewidth]{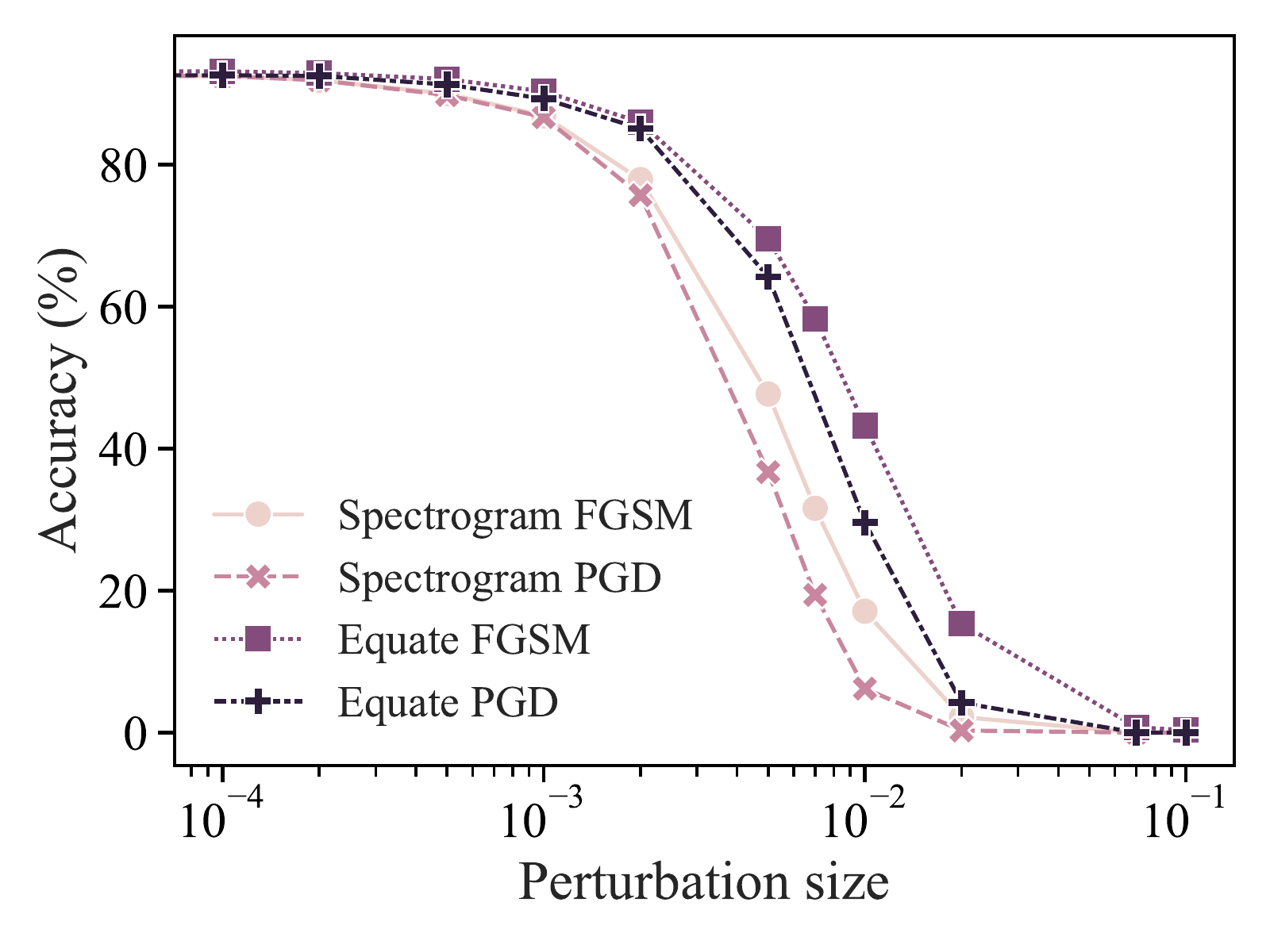}
\caption{ 
\textit{\small Symmetry constraint for spectograms.} \small Comparing FGSM and PGD attacks on spectorgrams for the Vggvox (\texttt{SBP}) without (Spectogram) and with (Equate) the symmetry constraint preserved. For a fixed accuracy degradation, more distortion and computation is needed to respect the Hermitian symmetry.
}
\label{fig:perturbation_spectro}
\vspace{-5mm}
\end{figure}

For each of the spectrograms, an adaptive adversary performs the naive attack in \S~\ref{subsec:constraints}, but ensures that symmetrical pairs of points in the spectrogram are the same. As presented in in Figure~\ref{fig:perturbation_spectro}, experimentally this attack is successful but the induced perturbation doubles. Importantly, observe that to maintain the same accuracy decrease the computationally expensive \textit{Equate PGD} attack (preserving the Hermitian symmetry) incurs higher distortion than the cheap standard \textit{Spectrogram FGSM} attack (oblivious to the symmetry constraint). Since FGSM is a PGD attack with 1 iteration, the cost to the adaptive adversary has been increased more than $100\times$.

\section{Imperceptible Attack: Kenansville Attack}
\label{app:sec:kenan}

\begin{figure}[t]
\centering
\includegraphics[scale=0.15]{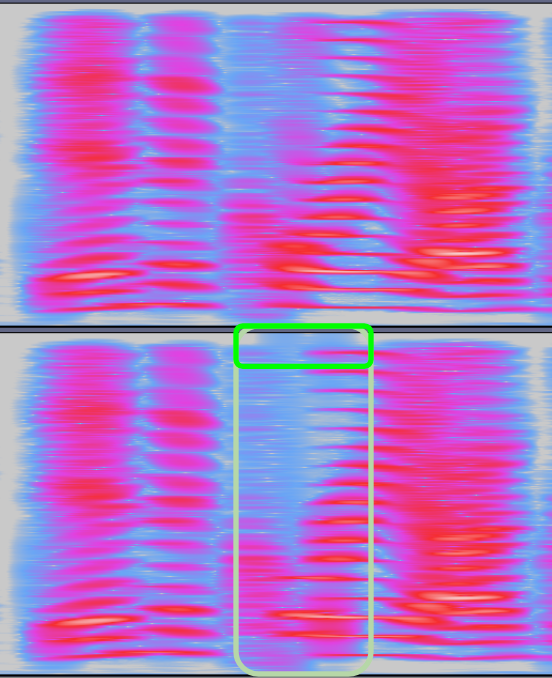}
\caption{ 
\textit{\small Spectrogram Comparison:}\small
\textbf{Top} - Original audio.
\textbf{Bottom} - Attacked audio. Note the removed spectrogram components (large box) (large box) and localized high frequency saturation (small box).}
\label{fig:kenan}
\vspace{-5mm}
\end{figure}
Observe that an ASI defender with a proactive approach can choose the segments of the speech that are uttered and analyzed. A defender may thus cherry-pick the segments and phonemes that are the hardest for the adversary to perturb. 

Consider the state of the art Kenansville attack \cite{abdullah_hear_2019}, a signal based attack on ASI performed by thresholding away spectral components humans would not perceived. In Figure \ref{fig:kenan}, we use the commodity graphical tool Audacity to display the spectrogram of a sample \texttt{wav} file provided on the paper's website. This check requires no domain specific skill from the defender. The attack changes the high frequencies in the signal not only by removing components of the speech but also by locally saturating the high frequencies. The saturated audio is not necessarily indicative of an attack itself. Yet combined with a strong suspicion of an attack, for instance unusual account activity, it can confirm that an attack has occurred. Here, the proactive defender may specifically choose vowels that are hard, \textit{i.e.} need to be entirely removed for the attack to succeed, forcing the more apparent spectral clues.

\section{Overt attack: frequency insertion}
\label{app:sec:frequency_insert}
Our adversary has simple and cheap tools to fool or reduce the accuracy of audio ML models as we show with a simple attack on \texttt{SPB}. We insert sine waves in the 20-4000Hz range into a signal, rendering the \texttt{SBP} system unusable. This blackbox attack requires no gradient optimization making its cost negligible compared to existing attacks. However, this attack is distinctly abnormal even with limited knowledge of audio attacks (see MTurk study in \S \ref{subsec:mturk}).

\section{Targeted SNES}
\label{app:sec:targeted}

\noindent{\bf Setup:} Our experimental setup is the same as in the untargeted case with some notable changes. We add an optimization attack based on the \textit{off-the-shelf} Adam optimizer~\cite{DBLP:journals/corr/KingmaB14}. Since we are adding an extra layer of complexity with the cross-pipeline transfer, we improve the similarity between our end-to-end surrogate and target by distillation~\cite{hinton_distilling_2015}. We implement an ensemble-based version of our SNES attack to boost targeted transferability~\cite{liu_delving_2017}. The models in the ensemble are obtained by distillation with a range of temperatures, both with and without a weighted sum of soft and hard labels~\cite{hinton_distilling_2015}. 

\noindent{\bf Results:} We randomly choose 1000 samples and use the samples that have the highest confidence vector for the \texttt{DBP} ensemble as target. We report only the best transfer rates for the adversary (100 rounds of Adam optimization). When combining Adam optimization, ensemble strategy and distillation, we generate adversarial examples that transfer for just 1.2\% of samples with top-1 accuracy and 4\% with top-5 match. The accuracy for true labels in that case is 21.9\%. Using the same hyperparameters, we then perform a target set attack\footnote{The goal is to obtain any one target in a set of targets.} using the 5 most likely but incorrect predictions of our surrogate as targets for each sample; our maximum transfer rate is 6.4\% with top-1 accuracy. The accuracy for the true labels of the model in that case is 14.7\%. Still, putting these last results in perspective the random prediction rate is 0.04\% (2484 classes), \textit{i.e.}, our top-1 rate is 160$\times$ better than random guess.

\end{document}